\documentclass[useAMS,usenatbib]{mn2e}

\usepackage{pict2e}
\usepackage{graphics}
\usepackage{rotating}
\usepackage{epsfig}
\usepackage{color}
\usepackage{multirow}
\usepackage{float}

\newlength{\colwidth}
\newcommand{\hMsol}{\, h^{-1}{\rm M_{\odot}}}
\newcommand{\hMpc}{\, h^{-1}{\rm Mpc}}
\newcommand{\hkpc}{\, h^{-1}{\rm kpc}}

\DeclareGraphicsExtensions{.eps}

\def\aj{AJ}					
\def\araa{ARA\&A}				
\def\apj{ApJ}					
\def\apjl{ApJL}					
\def\apjs{ApJS}					
\def\aap{A\&A}					
\def\mnras{MNRAS}				
\def\prd{Phys.~Rev.~D}				
\def\pasp{PASP}					
\def\nat{Nature}				

\title[Baryonic back-reaction]
{Impact of baryon physics on dark matter structures: a detailed simulation study of halo density profiles}

\author[A. R. Duffy et al.]
{Alan R. Duffy,$^{1,2,3}$ Joop Schaye,$^{2}$ Scott T. Kay,$^{1}$ Claudio Dalla Vecchia,$^{2,4}$ \newauthor Richard A. Battye$^{1}$ and C. M. Booth$^{2}$ \\
$^1$Jodrell Bank Centre for Astrophysics, School of Physics and Astronomy, The University of Manchester, Manchester M13 9PL, U.K.\\
$^2$Leiden Observatory, Leiden University, PO Box 9513, 2300 RA Leiden, The Netherlands \\
$^3$ICRAR, The University of Western Australia, M468, WA 6009, Australia \\
$^4$Max Planck Institute for Extraterrestrial Physics, Giessenbachstra§e, D-85748 Garching, Germany \\
}
\begin{document}

\date{\today}

\pagerange{\pageref{firstpage}--\pageref{lastpage}} \pubyear{2008}

\maketitle

\label{firstpage}

\begin{abstract}
The back-reaction of baryons on the dark matter halo density profile
is of great interest, not least because it is an important systematic
uncertainty when attempting to detect the dark matter. Here, we draw
on a large suite of high resolution cosmological hydrodynamical
simulations, to systematically investigate this process and its
dependence on the baryonic physics associated with galaxy formation.
The effects of baryons on the dark matter distribution are typically
not well described by adiabatic contraction models. In the inner ten per
cent of the virial radius the models are only successful if we allow
their parameters to vary with baryonic physics, halo mass and
redshift, thereby removing all predictive power. On larger scales the
profiles from dark matter only simulations consistently provide better
fits than adiabatic contraction models, even when we allow the
parameters of the latter models to vary.  The inclusion of baryons
results in significantly more concentrated density profiles if
radiative cooling is efficient and feedback is weak. The dark matter
halo concentration can in that case increase by as much as 30 (10) per
cent on galaxy (cluster) scales.  
The most significant
effects occur in galaxies at high redshift, where there is a strong
anti-correlation between the baryon fraction in the halo centre and
the inner slope of both the total and the dark matter density
profiles.  If feedback is weak, isothermal inner profiles form, in
agreement with observations of massive, early-type galaxies. However, we
find that AGN feedback, or extremely efficient feedback from massive
stars, is necessary to match observed stellar fractions in groups and
clusters, as well as to keep the maximum circular velocity similar to the virial velocity as
observed for disk galaxies. 
These strong feedback models reduce the baryon fraction in galaxies by a factor of 3 
relative to the case with no feedback. The AGN is even capable of reducing the baryon
fraction by a factor of 2 in the inner region of group and cluster haloes.
This in turn results in inner density profiles which are typically shallower than
isothermal and the halo concentrations tend to be \emph{lower} than in
the absence of baryons. We therefore conclude that the disagreement
between the concentrations inferred from observations of groups of
galaxies and predictions from simulations that was identified by Duffy
et al.\ (2008) is not alleviated by the inclusion of baryons.
\end{abstract}


\section{Introduction}
\label{sec:Introduction}
Increasingly powerful computers, combined with ever more 
efficient {\it N}-body codes, have permitted accurate numerical tests 
of structure formation in a Cold Dark Matter (CDM) universe
\citep[e.g.][]{Klypin:96, NFW, Moore:99, Bullock:01, Springel2005, VL2, Gao:08}.
It is now an established prediction that CDM haloes form a {\it cuspy} mass
distribution that is close to a `universal profile', independent of halo 
mass~\citep[][hereafter NFW]{NFW:96,NFW}. 
In detail, however, the haloes are not strictly self-similar~\citep{Navarro:10}.
Foremost amongst these results has been the existence of power law 
relationships in the phase-space density of haloes across several decades in 
radius~\citep{TaylorNavarro:01} and the halo concentration-mass 
relation over five decades in mass~\citep{Bullock:01,Neto:07,Maccio:08,Duffy:08b}.
High resolution {\it N}-body simulations have also shown that DM haloes are 
not smooth, but contain self-bound substructures that have survived the accretion
process~(e.g.~\citealt{MooreGhigna:99,Klypin:99}). More recently, it has been shown that
substructures themselves contain substructures,  
a pattern that shows no signs of abating~\citep{ghalo,aquarius}.

These power laws and embedded substructures are the product of an
approximately scale-free system that interacts only via gravity and, for
scales large enough that gravity is the only dynamically significant force,
can be viewed as an accurate approximation to structure formation. 
On smaller scales, other processes significantly modify the
structure of the baryons, which can subsequently affect the DM in the densest
regions. In particular, it has long been known that a galaxy forms because of the 
ability of its progenitor gas to cool efficiently via emission of radiation 
(e.g.~\citealt{Hoyle:53}). Such a process introduces a characteristic physical scale into the system
and thus breaks the self-similarity of structures~\citep{WhiteRees:78}.
The net effect is that the cooled gas and stars then cluster on smaller scales than the 
DM, thereby deepening the potential well of the system
and causing the DM to orbit closer to the centre. This will modify a number of $N$-body simulation 
predictions on small scales, in a way that depends on the efficiency of galaxy formation.

This modification of the DM distribution is usually parametrised in the form of adiabatic 
invariants, hence the term `adiabatic contraction' that is used 
to describe the overall effect, first considered by~\citet{Eggen:62} and \citet{Zeldovich:80}. 
The adiabatic parameter considered in the first generation of 
adiabatic contraction models for collapsing haloes~\citep{Blumenthal:86} was the product of 
radius with the mass internal to this radius, $M(<r)r$.
\citet{Gnedin:04}, henceforth G04, found that such a simple scheme, only
strictly applicable for particles on circular orbits, did not provide a good description
of what was happening in their hydrodynamic simulations of groups and clusters
at $z \approx 0$ and galaxies at $z > 3$. Better agreement was found
when they modified the parameter to use the orbit-averaged radius, $\bar{r}$, that
is a power-law function of radius. The work of G04 was extended to
galaxy scales at low redshift by~\citet{Gustafsson:06}, who modified the slope 
and normalisation of this power law distribution to provide a better description
of their results at this lower mass range.
The compression of the DM by the infalling baryons has also been extensively 
studied by purely analytic methods (e.g.~\citealt{Sellwood:05, Cardone:05, Klar:08}).

In addition to the contraction of the DM halo, the concentration of baryonic 
mass at its centre also leads to a more spherical 
structure~\citep{Kazantzidis:04}.
This effect is due to the more spherical gravitational potential formed by
the baryons and it is this that induces an adiabatic decrease in the 
eccentricity of the orbits of the DM and stars, rather than through the 
destruction of box orbits~\citep{Debattista:08}.
Additionally, in the presence of a gaseous disc, infalling subhaloes with low orbital 
inclinations with respect to the disc will experience significant disruption due to the
high local density of the baryons. This will lead to the accumulation of stars, and DM,
in the same plane, forming both a thin stellar and a DM disc~\citep{Read:08}.

With the addition of baryons to the simulation, one can expect further effects 
due to the transfer of angular momentum from infalling satellites
(e.g.~\citealt{Debattista:08,RomanoDiaz:08}). However,
the influence of baryons on these smaller, less well-resolved objects, 
are by no means certain. For example,
the substructure may exhibit greater resistance to tidal disruption due to 
its deeper potential well in the presence of baryons, while at the same time the
substructure will typically suffer from increased dynamical friction with the 
main halo~\citep{Jiang:08}.
Recent work by~\citet{Maccio:06} found a factor of two increase in the number of
surviving satellites in a galaxy-sized halo, in a hydrodynamical simulation
 relative to a DM-only simulation. 
This increased survival rate in the inner regions~\citep{Weinberg:08}
 may enable satellite infall to partly counteract the contraction of the 
DM halo, as these objects can efficiently transfer angular momentum to the inner
parts of the DM halo, instead of being tidally disrupted at larger radii
(for recent considerations of this effect see~\citealt{Pedrosa:09a},~\citealt{Pedrosa:10} 
and~\citealt{Abadi:09}). The erasure of the central DM cusp in galaxies by infalling
satellites was proposed by~\citet{ElZant:01} and for the case of clusters by~\citet{ElZant:04}.

The effect of the baryons on the inner DM density profile is of particular interest,
as it is currently a significant source of uncertainty when making predictions for 
DM detection experiments. For example, the expected $\gamma$-ray signal from 
self-annihilation of potential DM candidate particles~\citep{Springel:08},
which may become detectable in the near future using $\gamma$-ray observatories such as 
{\it Fermi}\footnotemark
\footnotetext{http://fermi.gsfc.nasa.gov}, is proportional to $\rho_{\rm DM}^2$.
We further note that the total mass density profile is also of interest, as it is
this quantity that is determined from strong lensing studies and is relevant 
for direct comparison with such observations (for a review 
of the methodologies and uses of strong lensing, see~\citealt{Kochanek:06}).

The incorporation of baryons in a simulation is a challenging theoretical 
undertaking as the relevant scales of the physical processes are rarely 
resolved in a cosmological simulation and approximations are therefore 
necessary. The method by which stellar feedback, for example, is incorporated
can have a large influence on the resulting baryonic distribution of the 
galaxy~\citep[e.g.][]{DallaVecchia:08}.
For this reason, we have attempted to probe these effects in a 
systematic way by examining the DM halo density profiles from galaxies, 
groups and clusters, drawn from an extensive series of cosmological simulations with 
that incorporate a wide variety of prescriptions for baryonic processes. 
These form a subset of the larger,
overall simulation series known as the OverWhelmingly Large
Simulations project 
(OWLS; \citealt{Schaye:10}).

Our work has a number of novel features that allow us to extend recent studies 
that have probed the dynamical response of the DM halo profile to the presence of 
baryons.
For example, while~\citet{Gustafsson:06} examined a small number of galaxy-sized
 haloes at $z=0$, we provide results on both group and cluster mass scales, with significantly
 better statistics.
G04 examined the DM profiles in a similar mass range to our own, albeit limited to higher 
redshift for galaxy scales, and with similar mass resolution. They demonstrate the significant
impact that metal enrichment and stellar feedback have on the
overall gas distribution, although they did not vary the models systematically.
The recent work of~\citet{Pedrosa:09a} investigated the effects of
physics on galaxy scales by simulating a single halo. Here we provide
a more statistically-robust set of 
results, due to the large number of haloes we have resolved. 
(We note that  a direct comparison with their $z=0$ galaxy is unfortunately not yet possible and 
must await further simulations).
In summary, the large dynamic range in mass, from dwarf galaxies to clusters,
across a wide variety of physics implementations and with
high statistical significance, is a significant advance in the study of baryons in DM haloes.

The rest of the paper is organised as follows.
In Section~\ref{sec:simulations} we describe the range of simulations used in this study, 
along with the differences between implementations of gas physics. 
In Section~\ref{sec:baryondist} we investigate the distribution of the baryons within these
haloes as a function of the physics implementation. In Section~\ref{sec:compobs} we compare our 
haloes with both strong gravitational lensing observations at high redshift and stellar mass 
fractions at the present day. We discuss the implications of our findings 
for structure formation in the CDM model, with the proviso that certain simulations
can fulfil some observations but none can be used at all mass scales.
The analysis of the haloes formed in these simulations, via
their density profiles, is discussed in Section~\ref{sec:profiles}. We then
attempt to parametrise the DM density profile with existing theoretical 
profiles, and the total matter profile by comparing non-parametric measures of halo 
concentrations, in Section~\ref{sec:fitprofiles}.
In Section~\ref{sec:AC} we test whether we can mimic the influence of the baryons by 
making use of the adiabatic contraction models of~\citet{Blumenthal:86}  and G04.
Finally, we summarise our work in Section~\ref{sec:conclusion}.

\section{Simulations}
\label{sec:simulations}
This work is based on OWLS; a series of high-resolution simulations of cosmological
volumes with differing subgrid physics implementations \citep{Schaye:10}.
These simulations were run using a modified version of {\sc gadget-3}, itself a modified 
version of the publicly-available {\sc gadget-2} code~\citep{Springel2005b}. {\sc gadget} 
calculates gravitational forces using the Particle-Mesh algorithm 
(e.g.~\citealt{KlypinShandarin:83}) on large scales, supplemented by the
hierarchical tree method~\citep{BarnesHut:86} on smaller scales. Hydrodynamic
forces are computed using the Smooth Particle Hydrodynamics (SPH) 
formalism~\citep{SPH85,SpringelHernquist:02}.

The production simulations available in OWLS model cubes of 
comoving length $25$  and $100  \, h^{-1}{\rm Mpc}$ with $512^3$ gas and 
$512^3$ DM particles. For all runs, glass-like cosmological initial conditions were 
generated at  $z=127$ using the Zel'dovich approximation and a transfer function
generated using {\sc cmbfast} (v.~4.1; \citealt{CMBFAST}). 
Cosmological parameters were set to the best-fit values from the 3rd year
{\it Wilkinson Microwave Anisotropy Probe} data~\citep{WMAP3}, 
henceforth known as WMAP3. Specifically, the values for
[$\Omega_{\rm m},$ $\Omega_{\rm b},$ $\Omega_{\Lambda},$ $h,$ $\sigma_{8},$ $n_{\rm s}]$
were set to [0.238, 0.0418, 0.762, 0.73, 0.74, 0.95]. The primordial
mass fraction of He was set to 0.248. All runs that used the same box
size had identical initial conditions.

We have analysed a subset of 6 OWLS runs for each box size: a DM-only run 
(hereafter labelled DMONLY) and 5 additional runs that also follow the baryonic component with hydrodynamics, gas cooling and star formation. 
For DMONLY, the particle masses are $m=7.7\times 10^{6}$ and
$4.9\times 10^{8}\hMsol$ for the $25$ and $100\hMpc$ runs, respectively. For
the runs with baryons, each particle was split in two in the initial conditions, with the 
mass shared according to
the universal baryon fraction, $f^{\rm univ}_{\rm b} = \Omega_{\rm b}/\Omega_{\rm m} = 0.176$,
such that the DM (gas) particle mass is $6.3\, (1.4)\, \times 10^{6}$ and
$4.1\, (0.86)\, \times 10^{8} \hMsol$ for the $25$ and $100\hMpc$ runs, respectively.
The $25\hMpc$ simulations were run to $z=2$ while the $100\hMpc$ simulations were run 
all the way to $z=0$ (our results will focus on data from these redshifts). 
At early times the softening length was held constant in comoving
co-ordinates at 1/25 of the initial mean inter-particle spacing. Below $z=2.91$ the softening length was held fixed in proper units (thus,
the $z=0$ values were $0.5\hkpc$ and $2\hkpc$ respectively).

\subsection{Baryonic physics}

\begin{table*}
\caption{A list of all simulations, the corresponding names in the
  OWLS project \citep{Schaye:10} and the differences in the included
  subgrid physics.}
\centering
\begin{tabular}{llccc} \hline
Simulation & OWLS name & Cooling & Supernova & AGN \\
 & & & Feedback & Feedback \\
\hline
PrimC\_NFB & NOSN\_NOZCOOL & Primordial & None & None \\ 
PrimC\_WFB & NOZCOOL & Primordial & Weak & None \\ 
ZC\_WFB & REF & Metal & Weak & None \\ 
ZC\_SFB & WDENS & Metal & Strong & None \\ 
ZC\_WFB\_AGN & AGN & Metal & Weak & Yes \\ 
\hline
\end{tabular}
\label{tab:simlist}
\end{table*} 

For our purposes two main aspects of the baryonic physics have been varied: 
radiative cooling (with and without metal lines) and feedback (from 
supernovae and accreting supermassive black holes). The modification of the
baryon distribution will be primarily determined by these two competing effects. 
Within OWLS other physics prescriptions have been varied, 
such as the choice of stellar initial mass function for example, however these
will at most have a secondary impact on the baryon distribution. 
We use `PrimC' and `ZC' to label two 
different approaches to cooling, denoting the absence- or inclusion of 
metals to the cooling rate respectively. For the supernova feedback, we use `NFB', `WFB' and 
`SFB' to identify whether stellar feedback is absent, weak or 
strong. In addition, we have analysed a simulation that includes feedback due to accreting
supermassive black holes, associated with Active Galactic Nuclei; this run is 
labelled `AGN'. Table~\ref{tab:simlist} lists the runs with baryonic
physics, along with their original OWLS names \citep{Schaye:10}, and highlights
their key differences. We only provide brief descriptions here, for
further details please see \citet{Schaye:10} and the references given below.

In all the simulations with baryons, radiative cooling (and heating) rates are
implemented element-by-element using tables generated with the {\sc cloudy} radiative 
transfer code~\citep{Cloudy}, following the method described in~\citet{Wiersma:09a}.
We assume collisional equilibrium before reionisation ($z > 9$) and photoionisaton equilibrium 
afterwards, in the presence of an evolving UV/X-ray 
background~\citep{HaardtMadau:01}. In the `PrimC' runs cooling rates
were computed assuming primordial abundances. In those labelled `ZC' we also track line emission from nine metals:
C, N, O, Ne, Mg, Si, S, Ca and Fe. Both types of simulation also include cooling 
via free-free Bremsstrahlung emission and Compton cooling due to interactions between
the gas and the cosmic microwave background. For full details of the implementation of gas cooling in the simulations, see~\citet{Wiersma:09a}.

When gas is sufficiently dense it is expected to be multiphase and unstable to star formation
\citep{Schaye:04}. Our simulations lack both the resolution and the
physics to model the cold interstellar gas phase and we therefore
impose an effective equation of state, $P \propto \rho^{\gamma_{\rm eff}}$,
for densities that exceed our star formation threshold of $n_{\rm
  H}>0.1\,{\rm cm}^{-3}$. The normalisation is fixed such that $P/k =
1.08\times 10^{3}\,{\rm K}\,{\rm cm}^{-3}$ at the 
star formation density threshold. The index is set to $\gamma_{\rm eff} = 4/3$, 
which has the advantage that both the Jeans mass and the ratio of the Jeans length to the SPH 
smoothing length are independent of density, thus suppressing spurious
fragmentation \citep{Schaye:08}. \citet{Schaye:08} showed how the
observed Kennicutt-Schmidt law, $\dot{\Sigma}_\ast = A (\Sigma_{\rm
  g}/1~{\rm M}_\odot\,{\rm pc}^{-2})^n$ can be analytically converted into a
pressure law, which can be implemented directly into the
simulations. This has the advantage that the parameters are
observables and that the simulations reproduce the input star
formation law irrespective of the assumed equation of state. We 
use the \citet{Schaye:08} method, setting $A=1.515\times 10^{-4}\,{\rm
  M}_\odot\,{\rm yr}^{-1}\,{\rm kpc}^{-2}$ and $n=1.4$
\citep[][note that we renormalised the observed relation for a
  Chabrier IMF]{Kennicutt:98a}.

These star particles are assumed to be simple stellar populations with an initial mass function 
given by~\citet{Chabrier:03}. The metal abundances of the stellar
particles are inherited from their parent
gas particles. The stellar evolution, and the subsequent release of metals from the star particle,
depends on this metallicity (\citealt{Portinari:98};~\citealt{Marigo:01};~\citealt{Thielemann:03}).
The delayed release of 11 individual elements (the ones used for the
calculation of the cooling rates) by massive stars, from Type Ia and Type II supernovae, as well as
AGB stars, is tracked. Every time step they are shared amongst neighbouring gas particles 
according to the SPH interpolation scheme. For further details please 
see~\citet{Wiersma:09b}.

Stars formed in the simulations labelled `WFB' and `SFB' inject energy into local gas particles. 
This energy is in kinetic form, i.e.\ nearby gas
particles are kicked away from the stars. On average, each newly
formed star particle kicks $\eta$ times its own mass, where $\eta$ is
the dimensionless mass loading parameter, by adding a randomly
oriented velocity $v_{\rm w}$ to the velocity of each kicked
particle. The simulations labelled `WFB' use $\eta=2$ and $v_{\rm
  w}=600~{\rm km\,s^{-1}}$, which 
  corresponds to forty per cent of the SN energy for our initial mass
  function. Further details
of the feedback prescription can be found in~\citet{DallaVecchia:08}. Note that although we have
labelled the supernova feedback in this simulation ``weak'', it is in fact strong compared to 
many prescriptions in the literature.

The runs labelled `SFB' implement wind velocity and mass-loading parameters that
depend on the local density of the gas, $v_{\rm w} = 600~{\rm
  km}\,{\rm s}^{-1} \left (n_{\rm H}/0.1~{\rm cm}^{-3}\right )^{1/6}$ and 
$\eta =2 \left (v_{\rm w}/600~{\rm km}\,{\rm s}^{-1}\right )^{-2}$,
such that in the densest regions winds are launched 
with the largest speeds and smallest mass-loadings. For gas that follows the imposed
effective equation of state, this implies a wind speed that is proportional to the 
effective local sound speed, $v_{\rm w} \propto c_{\rm s, eff} \propto (P/\rho)^{1/2}$.
The result is that winds in the `SFB' model remove gas 
from higher mass systems more efficiently than in the `WFB' model.  
Note that the WFB and SFB models use the same amount of SN
energy. The difference in the efficiency thus results purely from
energy is distributed between mass-loading and wind velocity.

Finally, in the simulation labelled `AGN' supermassive black holes (BHs) are grown,
with subsequent feedback, at the centres of all massive haloes
according to the methodology of~\citet{Booth:09},
which is a substantially modified version of that
introduced by~\citet{Springel:05}. 
Seed BHs of mass $m_{\rm seed}=9\times 10^4\,{\rm M}_\odot$
(i.e.\ $10^{-3}\,m_{\rm g}$ in the 100$~h^{-1}\,{\rm Mpc}$ box, where
$m_{\rm g}$ is the initial mass of the gas particles), are
placed into every DM halo more massive than $4\times 10^{10}\,{\rm
  M}_\odot$ (i.e.\ $10^2$ DM particles in the 100$~h^{-1}\,{\rm Mpc}$
box). Haloes are identified by regularly running a Friends-of-Friends (FOF)
group finder on-the-fly during the simulation.  After forming, BHs
grow by two processes: accretion of ambient gas and mergers. Gas
accretion occurs at the minimum of the Eddington rate and the
Bondi-Hoyle \citeyearpar{BondiHoyle:44} rate, where the
latter is
multiplied by $(n_{{\rm H}}/10^{-1}\,{\rm cm}^{-3})^2$ for
star-forming gas (i.e.\ $n_{\rm H} > 10^{-1}\,{\rm cm}^{-3}$) to compensate for
the fact that our effective equation of state strongly underestimates
the accretion rate if a cold gas phase is present. Gas accretion
increases the BH masses as $\dot{m}_{\rm BH} =
(1-\epsilon_r)\dot{m}_{\rm accr}$, where $\epsilon_{\rm r} = 0.1$ is the
assumed radiative efficiency. We assume that 
15 per cent of the radiated energy (and thus 1.5 per cent of
the accreted rest mass energy) is coupled thermally to
the surrounding medium. \citet{Booth:09} showed that this model reproduces the
redshift zero cosmic BH density as well as the observed relations
between BH mass and both the stellar mass and the central stellar
velocity dispersion. 

\subsection{Halo definitions and density profiles}
\label{subsec:halo}

The principal quantity that is extracted from each simulation is the spherically-averaged
halo density profile, dominated in all but the central regions by the dark matter. Haloes
were first identified using the FOF algorithm, that links dark matter
particles using a dimensionless linking-length, $b=0.2$~~\citep{Davis85}. We then used
the {\sc subfind} algorithm~\citep{subfind,Dolag:09} to decompose the FOF group into separate, 
self-bound substructures, including the main halo itself. Finally, a sphere was placed at
the centre of each halo, where the centre was identified with the location of the minimum potential
particle in the main halo. The spheres were then grown until a specified mean internal 
density contrast was reached. 

A halo thus consists of all mass, $M_{\Delta}$, within the radius, $R_{\Delta}$, for which 
\begin{equation}
M_{\Delta} = \frac{4}{3} \pi R_{\Delta}^3 \, \Delta \, \rho_{\rm crit}(z)\,,
\end{equation}
where the mean internal density is $\Delta$ times the critical density,
$\rho_{\rm crit}(z)=3H(z)^2/8\pi G$.
If $M_{\Delta} = M_{\rm vir}$ and $R_{\Delta}=R_{\rm vir}$, then one can compute 
$\Delta$ from the spherical top-hat collapse model. In this case, we adopt the fitting
formula from \citet{BryanNorman98}, that depends on both cosmology and redshift 
\begin{equation}
\Delta = 18\pi^2 + 82x - 39x^2\,,
\end{equation}
where $x = \Omega_{\rm m}(z) - 1$,
\begin{equation}
\Omega_{\rm m}(z) = \Omega_{\rm m}(1+z)^3 \left( \frac{H_{0}}{H(z)}\right)^2\,,
\end{equation}
and $H(z)$ is the usual Hubble parameter for a flat universe
\begin{equation}
H(z) = H_{0} \, \sqrt{\Omega_{\rm m}(1+z)^3 + \Omega_{\Lambda}}.
\end{equation}
In our cosmology at $z=0$ ($2$) $\Delta=92.5$ ($168.5$). 
We will also consider overdensities that are integer multiples of the critical
density; $\Delta = 500$ and $2500$, with radii and masses denoted in the
standard way, e.g.\ $R_{500}$ and $M_{500}$ for $\Delta = 500$.   

Following ~\citet{Neto:07}, only haloes with more than $10^4$ DM particles
within $R_{\rm vir}$ are 
initially selected. Density profiles are then computed using a similar method to our 
previous work on DM-only  haloes~\citep{Duffy:08b}. To briefly 
summarise, we take 32 uniform logarithmic shells of width $\Delta \log_{10}(r) = -0.078$,
in the range $-2.5 \le \log_{10}(r/{\rm R_{\rm vir}}) \le 0$. We then sum the 
mass within each bin for the three components (gas, stars and DM) and divide by 
the volume of the shell. We fit model density profiles by minimising 
the difference of the logarithmic densities between halo and model, 
assuming equal weighting (this was shown by~\citealt{Neto:07} to be a 
self-consistent weighting scheme).

A key issue is the range of radii over which we can trust that the profiles have converged 
numerically. To check this, we have performed a convergence study, 
the results of which are presented in Appendix~\ref{appendix:restest}. In summary, we 
show that the minimum radius proposed by~\citet{Power:03}, henceforth P03, 
is appropriate for our simulations, even for the ZC\_WFB
model (P03 defined their resolution limit on the two-body dynamical relaxation timescale in the 
context of DM-only simulations). All haloes used in 
this work satisfy the criterion $R_{\rm P03} < 0.05\,R_{\rm vir}$, 
so we adopt the latter value as the minimum radius for our density profiles. 

A selection of well-resolved haloes are singled out for investigation
of the slopes of their inner profiles
(Section~\ref{sec:innerprofile}). For these haloes, we place even
higher demands on the particle number and ensure that haloes satisfy $R_{\rm P03} < 0.025\,R_{\rm vir}$ and use the latter value as the minimum radius in this instance. In practice this criterion is 
applied to the DMONLY simulation to define a lower (virial) mass limit: at $z=0$ this is 
$3\times 10^{13} \hMsol$ and at $z=2$ it is $5\times 10^{11} \hMsol$. The other
runs are then matched\footnotemark to the DMONLY haloes  that make this cut. 
All haloes used to study the inner profile have at least $6 \times 10^{4}$ DM particles within 
$R_{\rm vir}$.
\footnotetext{Haloes are matched according to their shared DM particles; simulations with different 
physics of the same volume and resolution have identical initial conditions and particle 
ID lists. This allows us to match coincident groups of DM particles between runs.}

Between the two simulation sets, we have selected a total of 552 (204) haloes at 
$z=0$ (2) in DMONLY and of these, 67 (32) meet the more stringent resolution demands for the inner profile study. 
The difference in halo numbers in the different gas physics simulations 
is less than $10$ per cent, primarily due to low-mass haloes having fewer than 
$10^{4}$ DM particles and hence missing the initial cut.
In the subsequent discussion we will use the terms; 
{\it Dwarf Galaxy}, {\it Galaxy}, {\it Group}, and {\it Cluster}, to 
correspond to $M_{\rm vir} [h^{-1}{\rm M_{\odot}}]$ in the fixed ranges 
[$ < 10^{12}$,$10^{12} - 10^{13}$,$10^{13} - 10^{14}$,$ > 10^{14}$] respectively.

\section{Halo Baryon Distribution}
\label{sec:baryondist}

\begin{figure}
  \begin{center}
    \epsfig{figure=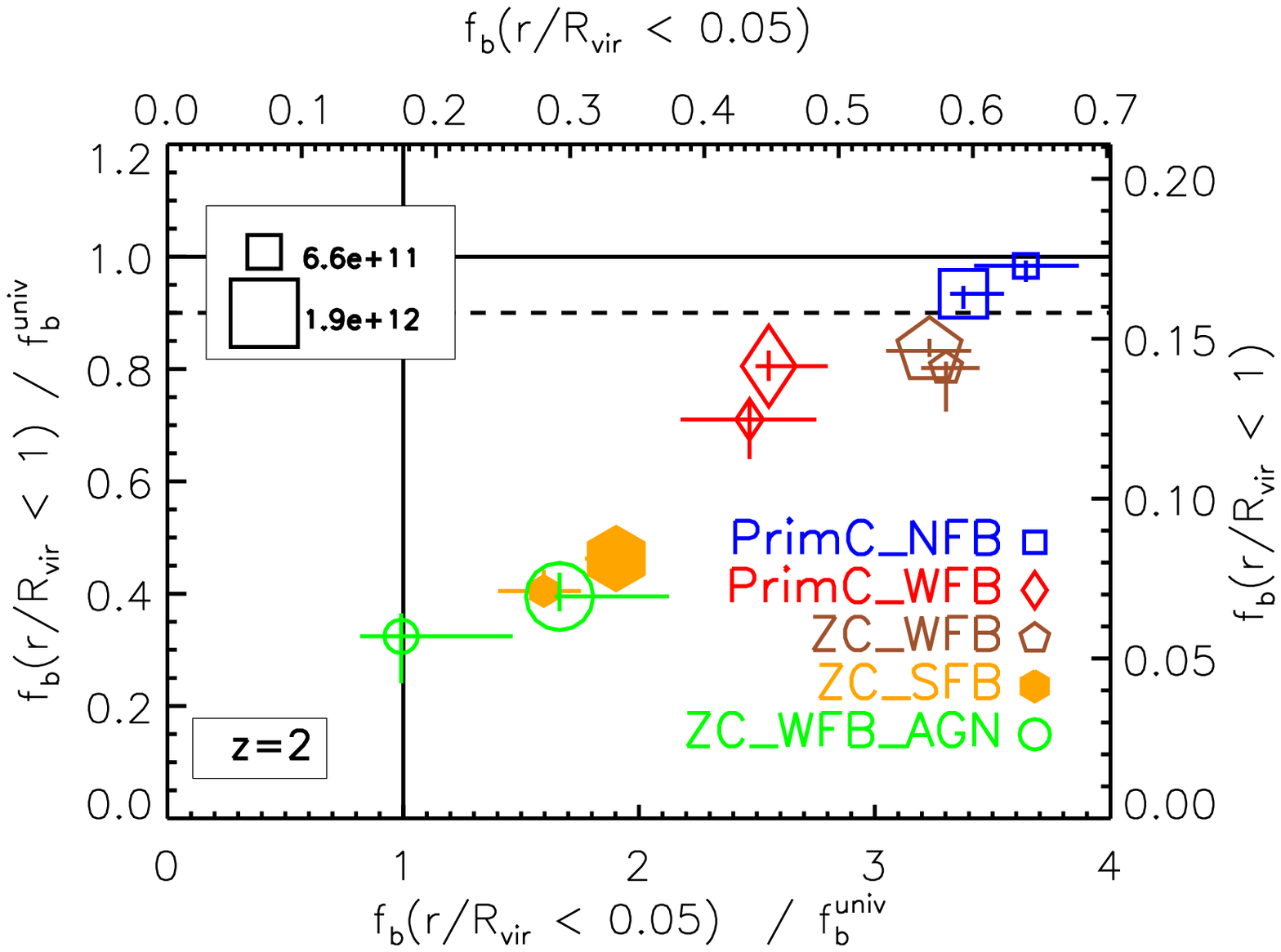, scale=0.45}
    \epsfig{figure=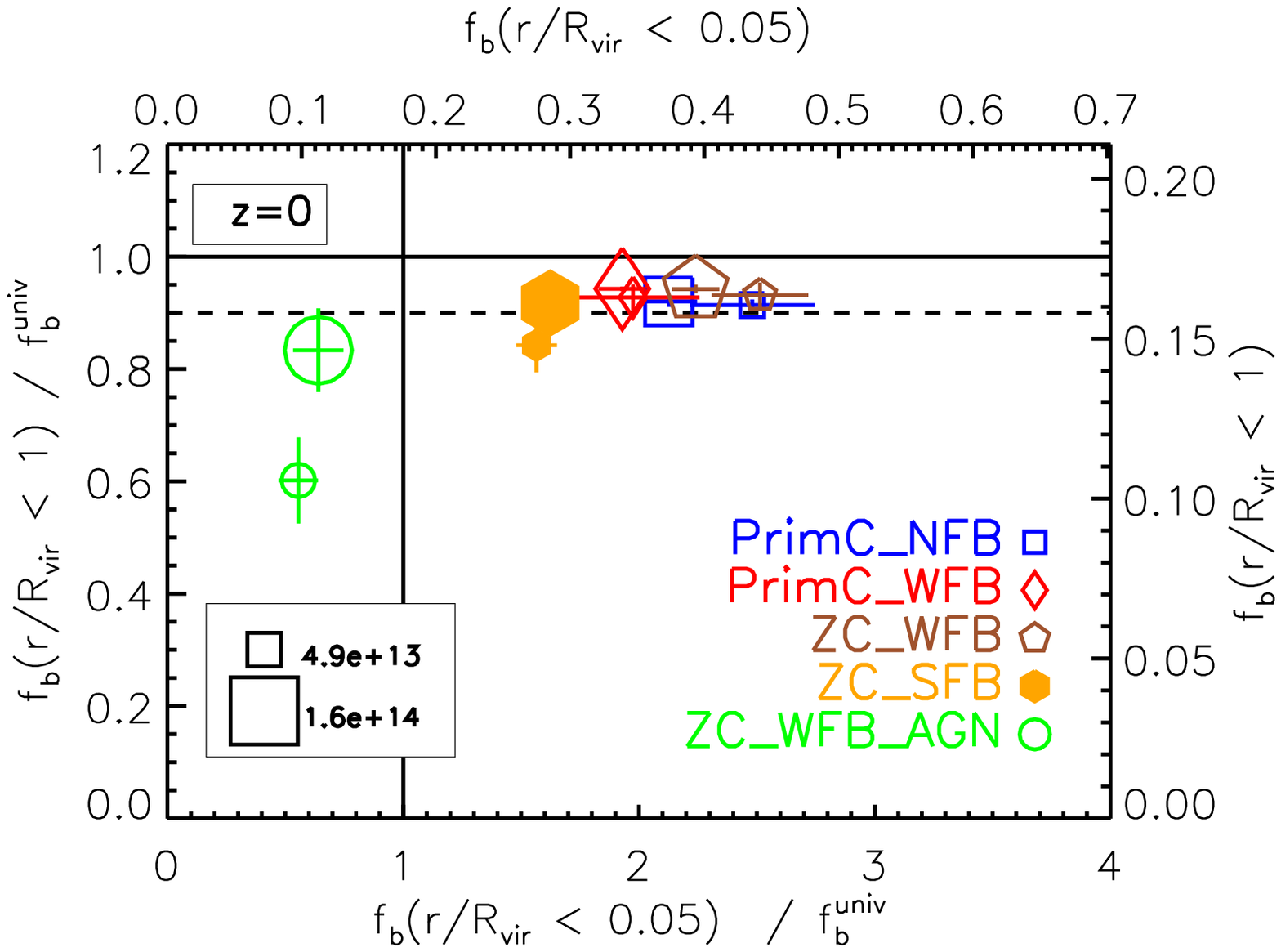, scale=0.45}
    \caption{The panels show the baryon fraction within the 
      virial radius as a function of the baryon fraction
      within $r < 0.05$ $R_{\rm vir}$, at $z=2$ (top panel) and $z=0$ 
      (bottom panel). The relative values of these 
      quantities gives a useful overview of the effect of cooling and feedback 
      on the baryon distribution in haloes at different redshifts and virial 
      masses. Each point corresponds to median values for haloes within 
      a given mass bin, indicated by the symbol size, 
      while the error bars indicate the quartile scatter.
      The solid vertical and horizontal lines 
      correspond to $f^{\rm univ}_{\rm b}$.
      The dashed horizontal line is from the non-radiative gas simulations 
      of~\citet{Crain:07}. {\it Galaxies} are more sensitive to the various feedback and 
      gas cooling schemes than {\it groups} and {\it clusters}.
      For {\it groups} and {\it clusters} the sensitivity to the feedback and cooling is mostly limited to 
      the inner regions but these processes affect {\it galaxies} all the way out to the virial radius. 
      AGN feedback reduces the baryon fractions most strongly.}
    \label{fig:fbfbvir}
  \end{center}
\end{figure}

The ability of baryons to cool radiatively and thereby cluster on smaller scales than the DM is
a recurring theme in this work, leading to, potentially, numerous modifications of the standard $N
$-body predictions. 
The exact distribution of the baryons in the halo will be sensitive to the physics 
implementation included and hence we should first quantify this key quantity. To that end we have 
plotted, in Fig.~\ref{fig:fbfbvir}, the baryon fraction, $f_{\rm b}$, inside the virial radius
against that within $r/R_{\rm vir}=0.05$.  We show the median value and the quartile scatter
of $f_{\rm b}$, within mass bins at the different redshifts. 
Only haloes that meet our more stringent {\it inner profile} criteria are shown.

A number of interesting features are apparent from the figure.
Firstly, we see that for {\it Groups} and {\it Clusters} at $z=0$ (bottom panel) the baryon 
fraction within the virial radius 
is close to the universal value in all runs except ZC\_WFB\_AGN. The other runs are
approximately consistent with, though slightly higher than, the prediction from non-radiative gas 
simulations~\citep{Crain:07}.
The deep gravitational potential wells and long cooling times in these objects, lead to little gas 
being expelled from the halo. Only in ZC\_WFB\_AGN, which includes AGN feedback and for 
which the feedback is therefore expected to be stronger on group and cluster scales, does a 
significant amount ($\sim 10-40$ per cent) of gas get expelled. It should be noted, however, that
a significant fraction of this gas will have been ejected from smaller systems at higher redshift.

The inner baryon fractions, $f_{\rm b}(r/R_{\rm vir} < 0.05)$,
typically decrease as the strength of the feedback increases.  
Note that ZC\_WFB\_AGN has an inner baryon fraction that is \emph{lower}
than the universal value. Runs with weak or no feedback have values
that are more than twice as high in the run with AGN feedback. 
The inclusion of metal-line cooling increases the inner baryon fraction, as can be seen 
by comparing ZC\_WFB with PrimC\_WFB.

The situation for {\it Galaxies} at $z=2$ (top panel) is very different.
Only the model with no supernova feedback, PrimC\_NFB, has a median baryon fraction 
within the virial radius that is close to $f_{\rm b}^{\rm univ}$; even the weak feedback
model is capable of expelling some gas. Note that PrimC\_NFB has a baryon fraction
that is higher than predicted from non-radiative simulations, suggesting that gas cooling
in this model has compensated for the heating and expansion of the gas due to energy transfer 
from the dark matter when the halo collapsed. In {\it Galaxy} mass haloes, the gravitational 
potential is sufficiently low that SN feedback is able to unbind gas from the halo. 
As a result, there is a strong positive correlation between  $f_{\rm b}(r/R_{\rm vir} <1)$ and 
$f_{\rm b}(r/R_{\rm vir}< 0.05)$. The galaxies in both ZC\_WFB and PrimC\_NFB have inner 
regions that are baryon-dominated ($f_{\rm b}>0.5$).

\section{Comparison with observations}
\label{sec:compobs}

In the previous section we compared simulations run with a number of widely used physical 
prescriptions, and demonstrated that  a large range of inner baryon fractions are potentially 
possible.   
We now determine which of these models, if any, are compatible with observations. We consider 
two probes that offer complementary and contrasting constraints on the baryon physics.  First, in 
Section \ref{sec:stellarfrac}, we present a comparison of the predicted redshift zero stellar mass 
fractions to those observed in {\it Galaxies}, {\it Groups} and {\it Clusters}.  
Secondly, in Section \ref{sec:SL}, we compare predicted total density profiles to those inferred for 
a high redshift {\it galaxy} sample through strong lensing measurements.  

\subsection{Stellar fractions}
\label{sec:stellarfrac}

\begin{figure}
  \epsfysize=2in \epsfxsize=4in
  \epsfig{figure=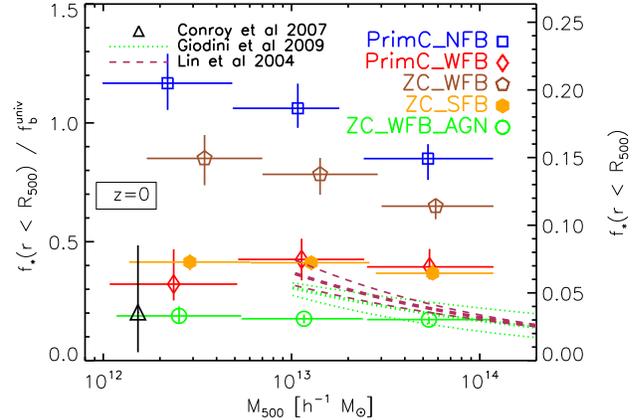, scale=0.45}
  \caption{To test the realism of our models we have compared the 
  median stellar mass fractions within $R_{500}$, in units of
    the universal baryon fraction, as a function of total mass within 
    $R_{500}$, at  $z=0$. 
    The coloured points represent the various physics prescriptions, apart from the
    single black data point at low
    halo mass which is from~\citet{Conroy:07} for a sample of isolated $L_{*}$ galaxies. 
    The vertical and
    horizontal error bars represent, respectively, the quartile spread and mass range within each bin. 
    The solid and dashed curves illustrate observational determinations from
    \citet{Giodini:09} and \citet{Lin:04}, with the outer curves representing 
    the uncertainty in the normalisation of their fits. It is clear that if the gas is allowed to cool by
    metal-line emission, it must be prevented from forming stars by either a supernova model that
    targets dense regions of the halo, ZC\_SFB, or by feedback from AGN, ZC\_WFB\_AGN.
    We note that although estimates of stellar masses are typically uncertain by a factor of 2-3
    (e.g. \citealt{Kupcu:07,Longhetti:09}) the observations still favour the strong feedback
    prescriptions.}
    \label{fig:fstar_m500}
\end{figure}

The stellar fraction by halo mass is well constrained observationally therefore a comparison 
of median stellar fraction (within $R_{500}$) to observation allows us to quickly determine the 
relative realism of our models. In Fig.~\ref{fig:fstar_m500} we show the stellar fraction, in units of 
$M_{500}$, for haloes at $z=0$ as a function of mass for our full range of physics models.

To compare with the observations, we have taken the group and cluster 
samples of~\citet{Lin:04} and~\citet{Giodini:09}. Note that the best fitting power law results we 
take from these works do not contain the contribution from the diffuse intracluster light, and can
hence be viewed as lower limits. This additional contribution likely ranges between 11 and 22
per cent (e.g. \citealt{Zibetti:05,Krick:07}). Following~\citet{Balogh:08}, we convert the observed
stellar luminosities from~\citet{Lin:04} to masses, assuming the best-fit stellar mass-to-light ratio,
$M/L_{K}=0.9$. Our simulations adopted a~\citet{Chabrier:03} IMF, 
whereas~\citet{Balogh:08} used the~\citet{Kroupa:01} IMF; we expect this difference
to be unimportant, however, as they are very similar over the relevant range of stellar masses.
For~\citet{Giodini:09} we have adopted the best fitting stellar fraction for the X-ray COSMOS
groups/poor clusters only. Also, we reduced the stellar mass fraction by 30 
per cent \citep{Longhetti:09} to account for their use of a~\citet{Salpeter:55} IMF.
Additionally, we include a result on {\it Galaxy} scales at $z=$ from~\citet{Conroy:07} who 
determined stellar masses from the spectroscopic SDSS value-added catalogue~\citep{Blanton:05},
along with halo mass estimates utilising the measured velocity dispersions of the satellites 
(we have scaled their halo mass, from $M_{200}$ to $M_{500}$, assuming an NFW profile with 
concentration given by~\citealt{Duffy:08b}). We further assume that the stars are significantly more 
concentrated than the DM and that modifying the mass cut from $R_{200}$ to $R_{500}$ will 
have a negligible effect on the stellar mass.
 We note that the result of~\citet{Conroy:07} is in agreement with~\citet{Mandelbaum:06} who 
 made use of weak lensing to determine halo masses.

The trend seen in our simulations, is a stellar fraction that decreases gently as $M_{500}$ 
increases from $10^{13}$ to $10^{14}\,{\rm M}_\odot$. This is in accord with the semi-analytic 
model results in~\citet{Balogh:08}.
Simulations with weak or no feedback contain stellar masses that are 2-3 times higher
than observed, PrimC\_WFB is an exception. This difference is expected as the dominant cooling 
channel at the virial temperatures of haloes with mass $\approx 10^{13}\,{\rm M}_\odot$ is 
believed to be from metal-line emission and hence `ZC' will have enhanced cooling rates over 
`PrimC'.
The simulations with stronger feedback, ZC\_SFB and ZC\_WFB\_AGN, are in better agreement 
with the observations.

We can see that simulations which have relatively efficient supernova feedback in the
dense regions of haloes, or include an additional heating source in the form of an AGN,
predict more realistic stellar fractions in present day {\it Groups} and {\it Clusters} with
$M_{500} \ge 10^{13} \,{\rm M}_\odot$. If we include the full mass range at low redshift, we can 
conclude that AGN feedback is necessary to prevent stellar mass building up in {\it Galaxies},
{\it Groups} and {\it Clusters}.
This is in good agreement with \citet{McCarthy:09}, who found that model ZC\_WFB\_AGN predicts
group K-band luminosities and gas fractions that are in excellent agreement with observations.
However, in the following section we will see that at high redshift the baryon-dominated inner halo 
of the schemes with weak or no feedback is required to match strong lensing results, 
leading to an intriguing conflict which we will discuss further in Section~\ref{sec:conclusion}.

\subsection{Observed inner profile slopes}
\label{sec:SL}

Gravitational lensing of light by an intervening galaxy enables us to
probe the inner mass profile of the lens galaxy's halo. The slope of the inner
mass profile for a sample of lens galaxies at $z<1$ was presented
by~\citet{Koopmans:06}. Surprisingly, the inner slope is strongly constrained to
be close to isothermal ($\beta \equiv \frac{d\ln\rho}{d\ln r} \approx -2$) with no evidence for
evolution. The region where the slope is measured
(the Einstein radius is typically around $3\hkpc$) is comparable to the scales
accessible in our highest-resolution simulations at $z=2$, allowing us to compare
the two results. 

\begin{figure}
  \epsfysize=2in \epsfxsize=4in
  \epsfig{figure=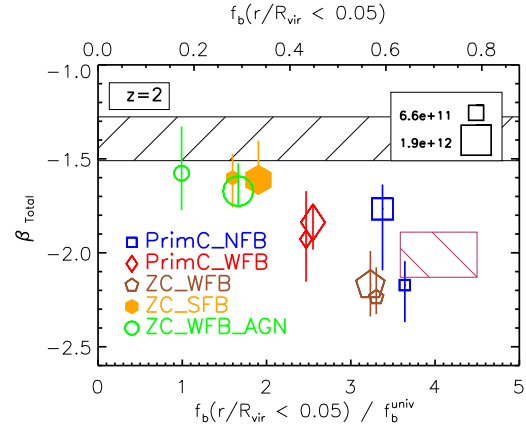, scale=0.45}
  \caption{Inner slope of the total mass density profile versus central baryon fraction.
  Only our results for {\it Galaxy} haloes at $z=2$ are 
    shown, to compare with the observational constraint on lensing galaxies 
    (\citealt{Koopmans:06};  the vertical size of the, lower, maroon box indicates 
    the intrinsic 1$\sigma$ Gaussian spread across all redshifts
    while the horizontal size of the box is the 68 per cent confidence interval of the 
    quoted stellar fractions, estimated by bootstrapping. Note that for the sample
    of early type galaxies the stellar fraction is essentially the baryon fraction). 
    The different symbol sizes denote different mass ranges. Symbol type and colour are
    used to distinguish different simulations. The black, hatched region indicates the quartile
    spread of the DMONLY simulation. As is clear from this figure, only
    simulations with high central baryon fractions reproduce the 
    observed steep inner density profiles of high redshift {\it Galaxies}. This is in contrast
    with the preferred simulation schemes at higher masses and lower redshift 
    (Fig.~\ref{fig:fstar_m500}).}
  \label{fig:totalbetafbprofile}
\end{figure}

\begin{figure}
  \epsfysize=2.25in \epsfxsize=4.5in
  \epsfig{figure=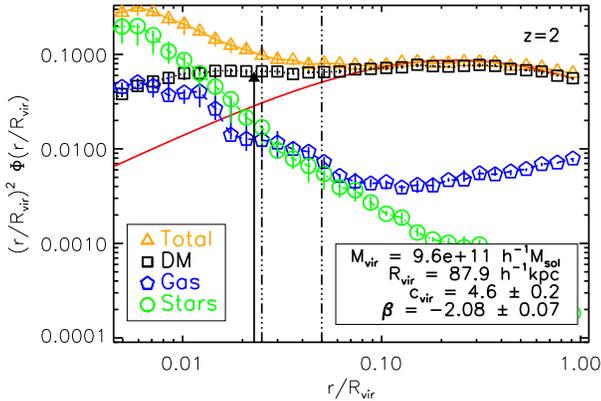, scale=0.45}
  \caption{We plot the mean density profile of a number of species, normalised 
    by the average virial mass and radius, and scaled by $r^2$ to reduce the 
    dynamic range. Here we can see the significant dynamical influence of the
    baryons in this particular object. This is a galaxy-sized object 
    at $z=2$ taken from ZC\_WFB, averaged over 20 relaxed haloes, from the 
    simulation with comoving length 25 $h^{-1}\,{\rm Mpc}$. The 
    yellow triangles are the total matter profile; the black squares are 
    the DM profile;
    the blue pentagons indicate the gas profile and the green circles denote 
    the stellar profile. Error bars illustrate the 68 per cent confidence 
    limits within each radial shell, estimated by bootstrap analysis. 
    Vertical lines indicate the region
    where the inner slope is measured ($0.025<r/R_{\rm vir} <0.05$). 
    The arrow represents $R_{\rm P03}$, taken to be the
    resolution limit. The solid red curve is the best-fit NFW profile to the DM, over the
    outer region ($r/R_{\rm vir} >0.05$). The DM mass was divided by 
    $(1-f_{\rm b}(r <R_{\rm vir}))$ such that the total mass of the DM component would equal $M_{\rm vir}$.
    With the halo mass known the NFW profile has only one free parameter. The NFW curve has had this
    factor removed for this plot. The legend 
    contains the mean virial mass and radius of the haloes,  $M_{\rm vir}$ and $R_{\rm vir}$
    respectively, the best-fit NFW concentration, $\rm c_{vir}$, and inner profile slope, $\beta$. 
    Baryons strongly influence both the total and the DM density profiles within $0.1\,R_{\rm vir}$.}
  \label{fig:allspeciesdensplot}
\end{figure}

Fig.~\ref{fig:totalbetafbprofile} shows the inner slope of the total mass density
profile versus central baryon fraction for our {\it Galaxy} haloes at $z=2$. 
When feedback is weak or absent, the inner slope is close to the 
isothermal value ($\beta=-2$); when feedback is strong, the slope is flatter
and close to the DMONLY values 
($\beta \sim -1.4$; shown as the black, hatched region, Section~\ref{sec:innerprofile}). Comparing these results with the observations of \citet{Koopmans:06}, shown as the lower
maroon box in the figure,
only the simulations with weak or no feedback produce similar values to the 
observations. The isothermal profile at $z=2$ is also seen in~\citet{RomanoDiaz:08}, who
found that for their simulations the influx of subhaloes at late times, $z \le 1$, acted to flatten 
the profile.

This conclusion is apparently at odds with what can be drawn from the observed
stellar fractions. On the one hand, strong feedback is required to keep cooling 
under control, such that the observed  stellar mass fractions in groups 
and clusters at low redshift are reproduced. On the other hand, the feedback 
has to be weak (or absent), to generate the steep inner profiles observed in 
galaxies at low and intermediate redshift (out to $z \sim 1$).  

One possible reason for this dichotomy, is that the simulated galaxies are at 
higher redshift than the lensing observations\footnote{We use the highest 
resolution simulation at $z=2$ due to the close match between typical Einstein 
radius of galaxies in~\citet{Koopmans:06}, $\approx 3\, h^{-1}\, {\rm kpc}$, and
our fitting range $\approx 2 - 4.5 \, h^{-1}\, {\rm kpc}$.}. However, as shown 
by~\citet{Dehnen:05}, the density profile of a collisionless system does not 
steepen during mergers, so the lensing galaxies must have been isothermal at
higher redshift, where it is more likely that significant gas condensation could
have occurred. Secondly, we cannot rule out selection effects in the observations that may
have a bias towards steeper density profiles. This could be important as isothermal profiles do 
exist in the ZC\_WFB\_AGN simulation, albeit significantly fewer in number than ZC\_WFB creates.
A final possibility is that the simulations themselves do not yet accurately 
model the high-redshift growth of the brightest galaxies. One would therefore 
require a feedback prescription that was less effective at high redshift 
(allowing more gas to accumulate and steepen the central profile). The feedback
must then rapidly expel the gas before it can form stars, faster than the 
dynamical timescale of the halo, such that the DM retains an isothermal profile.

\section{Halo Density Profiles}
\label{sec:profiles}

\begin{figure*}
  \begin{center}
    \begin{tabular}{cc}
      \epsfig{figure=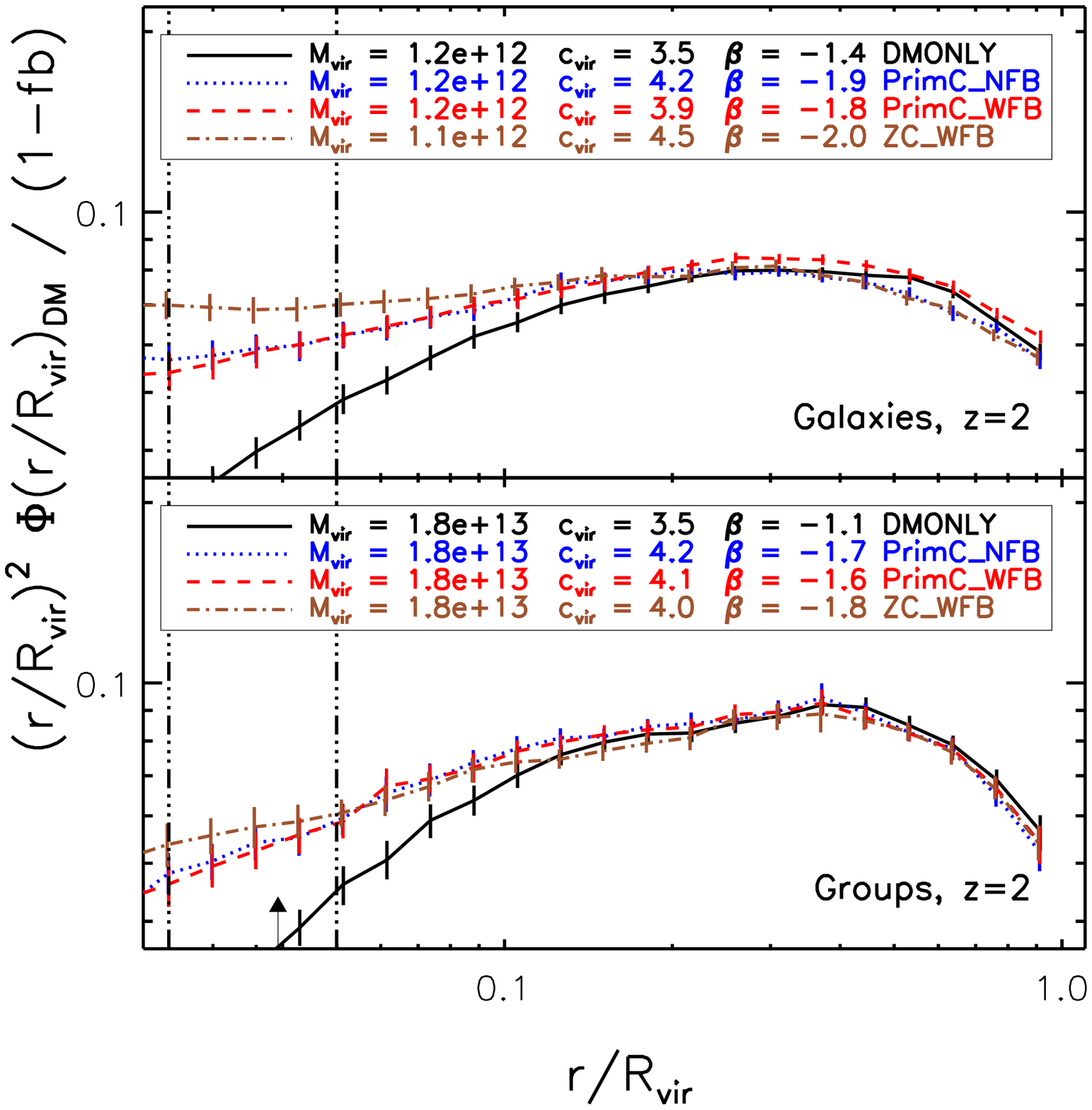, scale=0.45} &
      \epsfig{figure=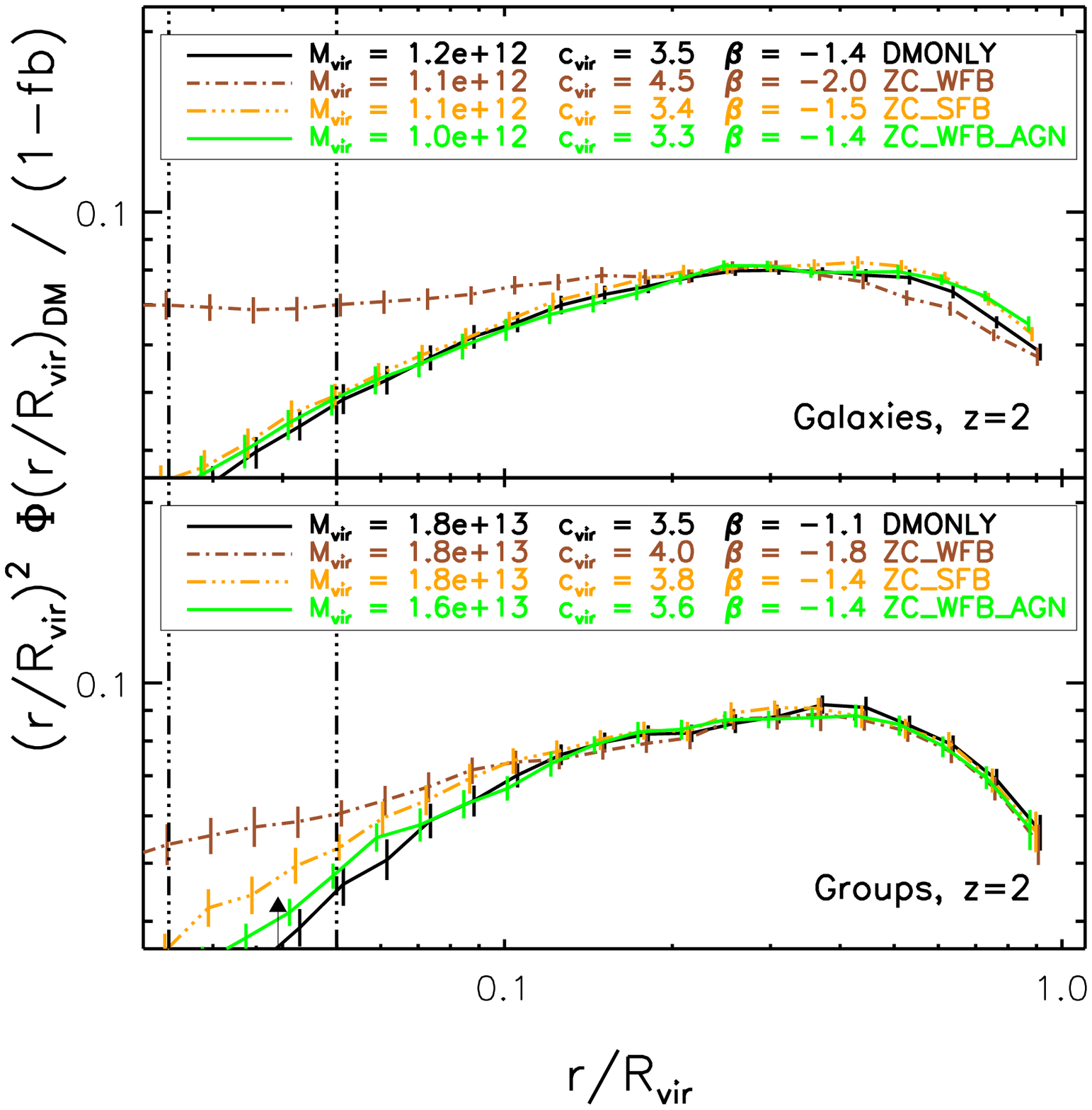, scale=0.45} \\
    \end{tabular}
    \caption{In order to quantify the response of the DM halo to the varying physics prescriptions, 
    we plot the mean DM halo density profile of haloes in fixed mass bins,
    $M_{\rm vir} = 5\times10^{11} - 5\times10^{12} h^{-1}\,{\rm M_{\odot}}$, the mean mass of which 
    corresponds to {\it Galaxies} (top panel) and
    $M_{\rm vir} = 10^{13} - 5\times10^{13} h^{-1}\,{\rm M_{\odot}}$ (the latter limit is to include the
    largest object)
    which are {\it Groups} (bottom) at $z=2$ (drawn from the 25 $h^{-1}\,{\rm Mpc}$ simulations).  
    Units are 
    normalised by the virial mass and radius from the 
    equivalent haloes in the DMONLY simulation. The DM density profiles for the models including
    baryons have been divided by $(1-f_{\rm b}^{\rm univ})$ to facilitate comparison with DMONLY.
    The left panels show results for the
    runs with weak or no feedback and the right panels for different feedback schemes. The 
    vertical lines denote $r/R_{\rm vir} =0.025$ and $0.05$, the region where the inner profile slope
    is estimated. The vertical arrows denote the P03 resolution limit; in the case of the 
    {\it Groups}, the inner profile slope cannot be reliably measured at $z=2$ (the other
    haloes have a resolution limit within the innermost line). Error bars are bootstrap 
    estimates of the 68 per cent confidence limits about the mean density within each bin. The 
    legend contains the mean virial mass of the haloes,  $M_{\rm vir}$, 
    the best-fit NFW concentration, $\rm c_{vir}$, and inner profile slope, $\beta$. }
    \label{fig:rhosqsims_z2}
  \end{center}
\end{figure*}

To first get a general idea of the effects of baryons on the DM halo, we show an example
of the halo density profile and its components in Fig.~\ref{fig:allspeciesdensplot}, where we have 
plotted the total mass profile, as well as the individual gas, stellar and DM components.
Here, the halo is an average over 20 relaxed {\it Galaxy} haloes from ZC\_WFB at $z=2$, and is 
shown in dimensionless form: $(r/R_{\rm vir})^2\Phi(r/R_{\rm vir})$, where 
$\Phi(r/R_{\rm vir}) \equiv \rho(r/R_{\rm vir}) R_{\rm vir}^3/M_{\rm vir}$. 

As can be seen in the figure, the baryons are more centrally concentrated than the DM. 
The stellar component has a steeper profile than the DM at all radii, whereas the gas
is steeper in the inner region ($r/R_{\rm vir} <0.05$) and slightly flatter in the outer region 
($r/R_{\rm vir} >0.1$). 
The inner region of the DM profile has been significantly affected by the baryons. 
To highlight this change, we also show the best-fit Navarro, Frenk \&
White (\citealt{NFW}; hereafter NFW) profile (solid, red curve), fit to data with $0.05 \le r/R_{\rm vir}  \le 1$. 
The NFW profile, which is a good approximation to the equivalent
DMONLY profiles, is of the form
\begin{eqnarray}\label{eqn:NFW}
\frac{\rho(r)}{\rho_{\rm crit}} = \frac{\delta_{\rm c}}{(r/r_{\rm s})(1+r/r_{\rm s})^{2}}\,,
\end{eqnarray}
where $\delta_{\rm c}$ is a characteristic density contrast and $r_{\rm s}$
is a scale radius. 
We fit for only one parameter, $r_{\rm s}$, as we use the previously-measured virial
radius and mass for each halo to define $\delta_{\rm c}$. 

The simulated DM profiles are significantly steeper in the inner regions than the NFW profile. 
This is expected:
the high central baryon fraction in this run (which, as we can see, is mainly in stars) has
pulled the DM in towards the centre. Measurement of the inner slope of
the DM profile ($\beta\approx-2$) confirms that the inner profile is isothermal. It is therefore clear 
that the effects of baryonic physics (radiative cooling in this case) can have a profound impact on 
the inner DM density profile. 

\subsection{Galaxies and groups at high redshift}
\label{sec:densityprofile_z2}

\begin{figure*}
  \begin{center}
    \begin{tabular}{cc}
    \epsfig{figure=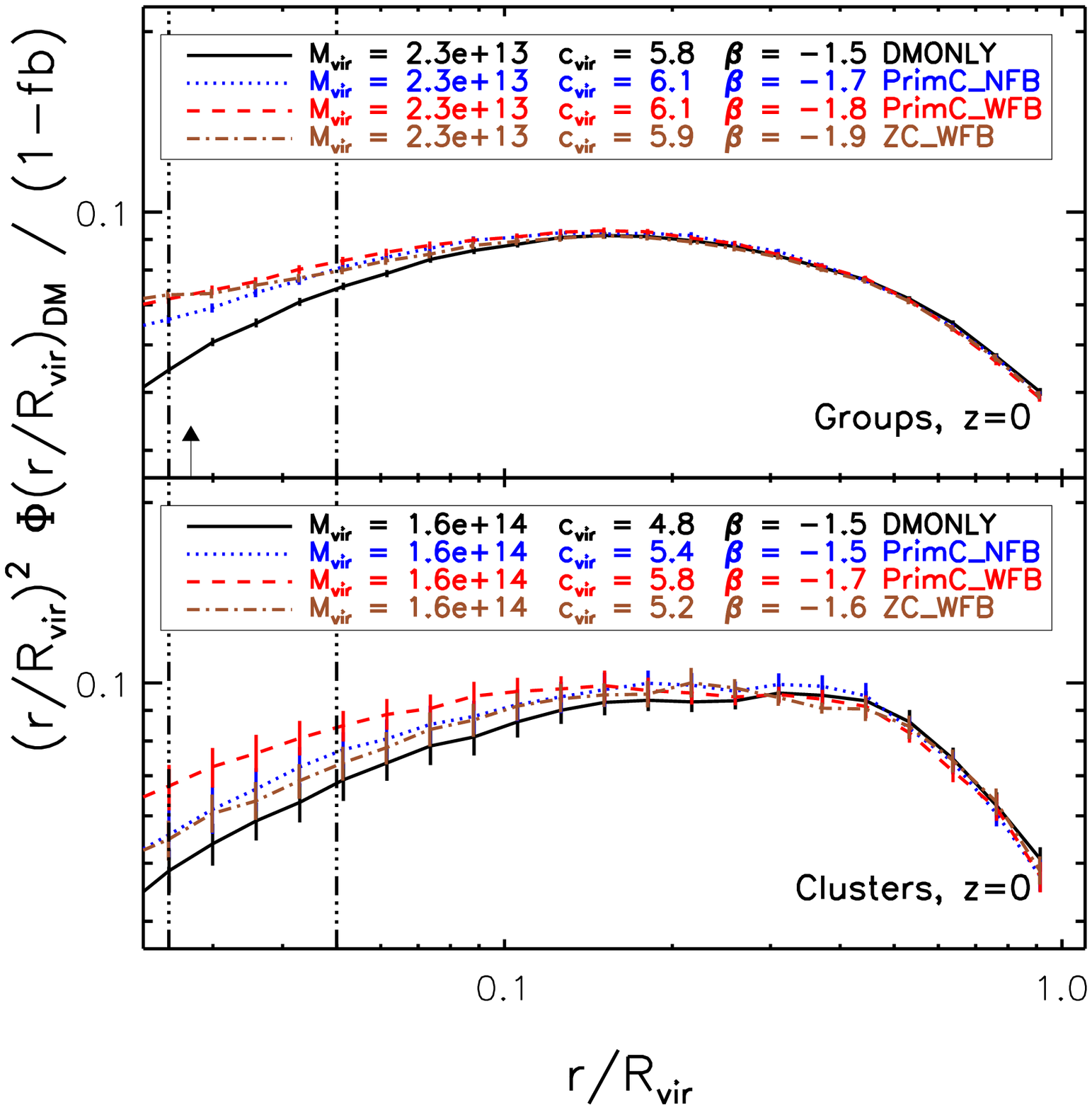, scale=0.45} &
    \epsfig{figure=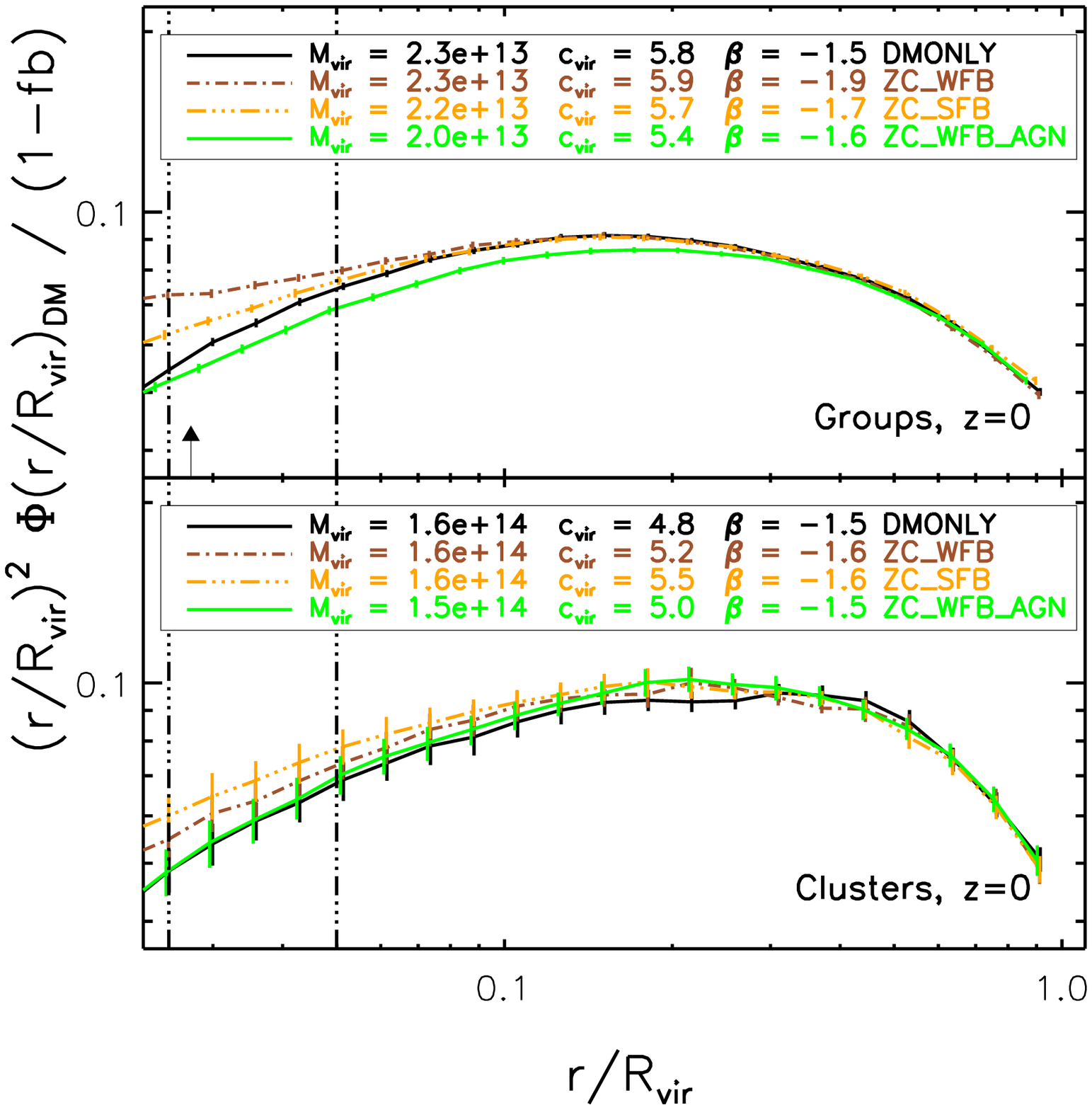, scale=0.45} \\
    \end{tabular}
    \caption{As in Fig.~\ref{fig:rhosqsims_z2} but now for {\it Groups} and {\it Clusters} at $z=0$ 
    (drawn from the $100\hMpc$ simulations). The top panels use a mass range 
    $M_{\rm vir} = 10^{13} - 10^{14} h^{-1}\,{\rm M_{\odot}}$ and the bottom panel for
    the range $M_{\rm vir} = 10^{14} - 6 \times 10^{14} h^{-1}\,{\rm M_{\odot}}$ 
    (the upper bound includes the most massive cluster in the simulation).}
    \label{fig:rhosqsims_z0}
  \end{center}
\end{figure*}

We present mean DM density profiles for the $z=2$ data in 
Fig.~\ref{fig:rhosqsims_z2}. These consist of 32 haloes from the 
$25 \,h^{-1}{\rm Mpc}$ simulation at {\it Dwarf Galaxy} and {\it Galaxy} mass scales (top panels) 
and an additional 14 {\it Group} scale objects (bottom panels) drawn from the larger $100\hMpc$
simulation.
Note that the group scale objects do not satisfy our stringent criterion for the inner profile 
study: the black arrow, depicting the P03 resolution limit, is at $r/R_{\rm vir} \approx 0.04$.

We have grouped the simulations together according to the physics 
prescriptions that we are testing. The first set is shown in the left panels and
corresponds to the introduction of metal-line cooling and
weak stellar feedback. The second group, in the right panels, tests the
types of feedback, namely stellar (both weak and strong) and AGN, all with the
same cooling prescription. The DM density profiles have been divided by
$(1-f^{\rm univ}_{\rm b})$ for comparison with the DMONLY profiles.

As expected, the largest differences occur in the inner regions, where the density is highest. In the 
left panels, showing results for runs in which feedback effects are weak or absent, 
there is a significant steepening of the density profile towards
higher densities on smaller scales across all mass ranges. 
The curves then more-or-less converge with DMONLY at radii larger than
$r/R_{\rm vir} = 0.2$. Given that PrimC\_WFB and PrimC\_NFB predict
nearly identical DM profiles, but differ strongly from DMONLY, it is clear that cooling
plays a crucial role in determining the magnitude of the back-reaction on the DM. 
Indeed, comparing PrimC\_WFB with ZC\_WFB, we see that including metal-line
cooling results in a steeper profile for {\it Galaxies}. This difference is reduced
in more massive systems for which the virial temperatures exceed the
regime in which metal-line cooling is efficient
\citep[e.g.][]{Wiersma:09a}. 

In the right panels of Fig.~\ref{fig:rhosqsims_z2} we see that the stronger feedback
schemes (ZC\_SFB and ZC\_WFB\_AGN) have DM density profiles that are closer to
those from DMONLY. These runs have smaller baryon fractions, thus the overall effect
of the cooling has been reduced. 

\subsection{Groups and clusters at the present day}
\label{sec:densityprofile_z0}

For {\it Groups} and {\it Clusters} at $z=0$ we repeat our previous
investigation 
in Fig.~\ref{fig:rhosqsims_z0}. Again, the DM profiles in the simulations with baryons
diverge from the DMONLY results in the inner regions, although the effect is not as 
dramatic as it was for the less massive systems shown in Fig.~\ref{fig:rhosqsims_z2}. This is because the typical cooling times are longer in these systems.
Furthermore, the difference between the runs with weak and strong stellar feedback (ZC\_WFB and ZC\_SFB; 
right panels) is smaller, because the strong stellar feedback is 
less effective in the more massive {\it Groups} and {\it Clusters} than it was in the lower-mass
systems at high redshift. The ZC\_WFB\_AGN model (with the  lowest central baryon 
fractions) produces almost identical profiles to DMONLY in {\it Clusters}, but a 
slightly \emph{lower} profile in {\it Groups}. 

Another interesting effect apparent in Fig.~\ref{fig:rhosqsims_z0}
for the {\it Clusters} at $z=0$ concerns the difference between the DM profiles of the 
no and weak feedback schemes (left panels). The model showing the
highest central DM density is PrimC\_WFB, while PrimC\_NFB and ZC\_WFB are statistically indistinguishable. At first glance this is unexpected,
as the latter two models have higher central baryon fractions (see
Fig.~\ref{fig:fbfbvir}, bottom panel). 
A similar result was found by~\citet{Pedrosa:09a} for their galaxy simulations at $z=0$. 
We summarise their explanation of the effect within the context of our simulations.
When weak stellar feedback is included (going from PrimC\_NFB to PrimC\_WFB),
a certain amount of the gas will be expelled from satellite galaxies but not from the
main (group or cluster) halo itself. As a result, the satellite haloes will be less bound
and suffer more tidal disruption as their orbit decays due to dynamical friction. The 
result is that less mass is transferred to the centre of the halo. When metal enrichment
is added to the simulation (going from PrimC\_WFB to ZC\_WFB), the cooling time
of the satellite gas becomes shorter (due to metal-line emission) and as a result, the 
satellite is able to hold on to more of its gas. This reduces the effect the feedback
has on the evolution of the satellite halo itself (and thus more mass can be transferred
to the centre).
The most important consequence of these effects is where the angular momentum
of the satellites gets transferred. In the case of PrimC\_WFB, less of the angular momentum 
is transferred to the inner region than in PrimC\_NFB, for example. As a result, the inner
profile is denser in the former case, even though the overall baryonic mass is smaller
in the centre of the halo.

\subsection{Inner profiles}
\label{sec:innerprofile}
\begin{figure}
  \epsfysize=2in \epsfxsize=4in
  \epsfig{figure=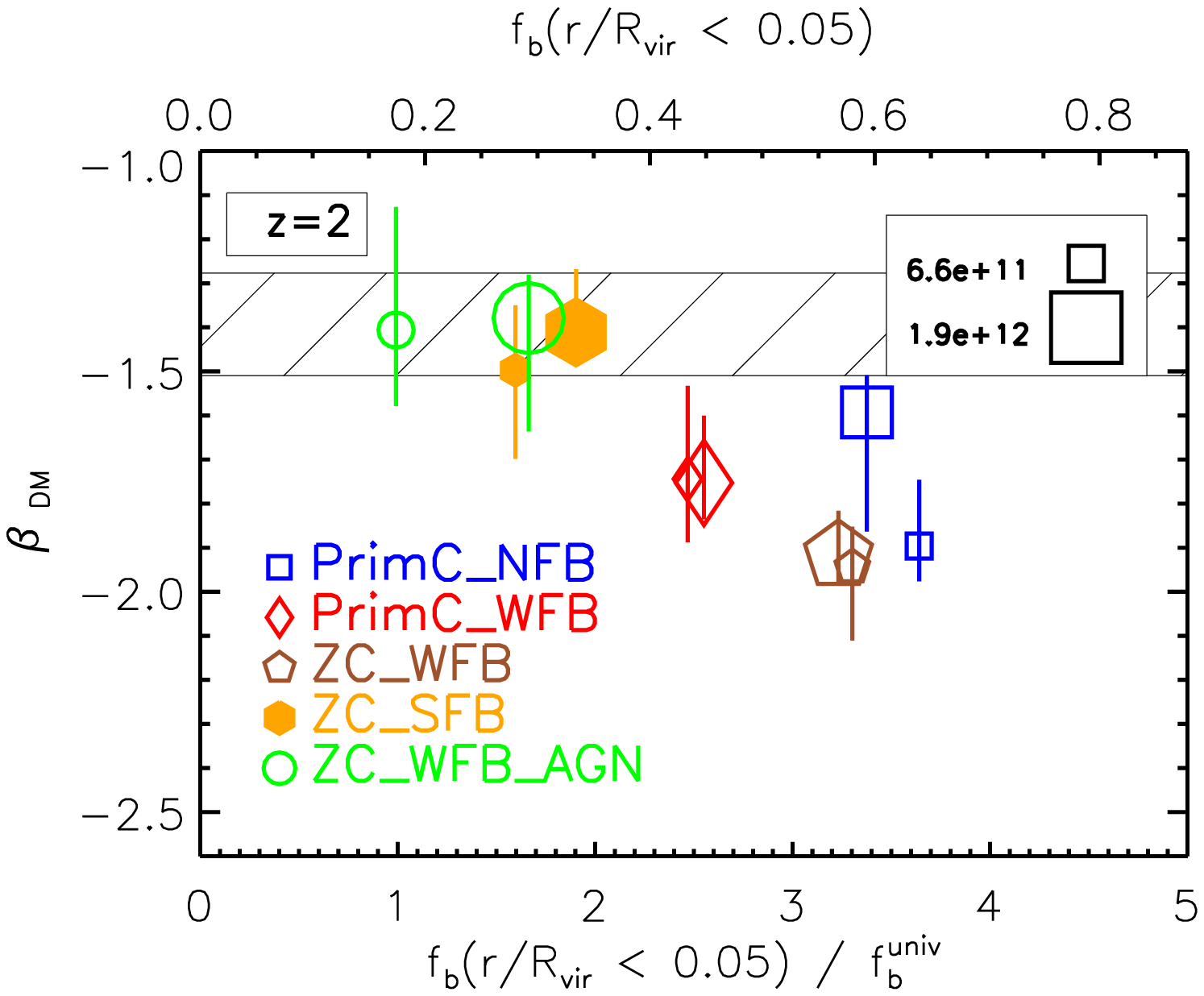, scale=0.45}
  \epsfig{figure=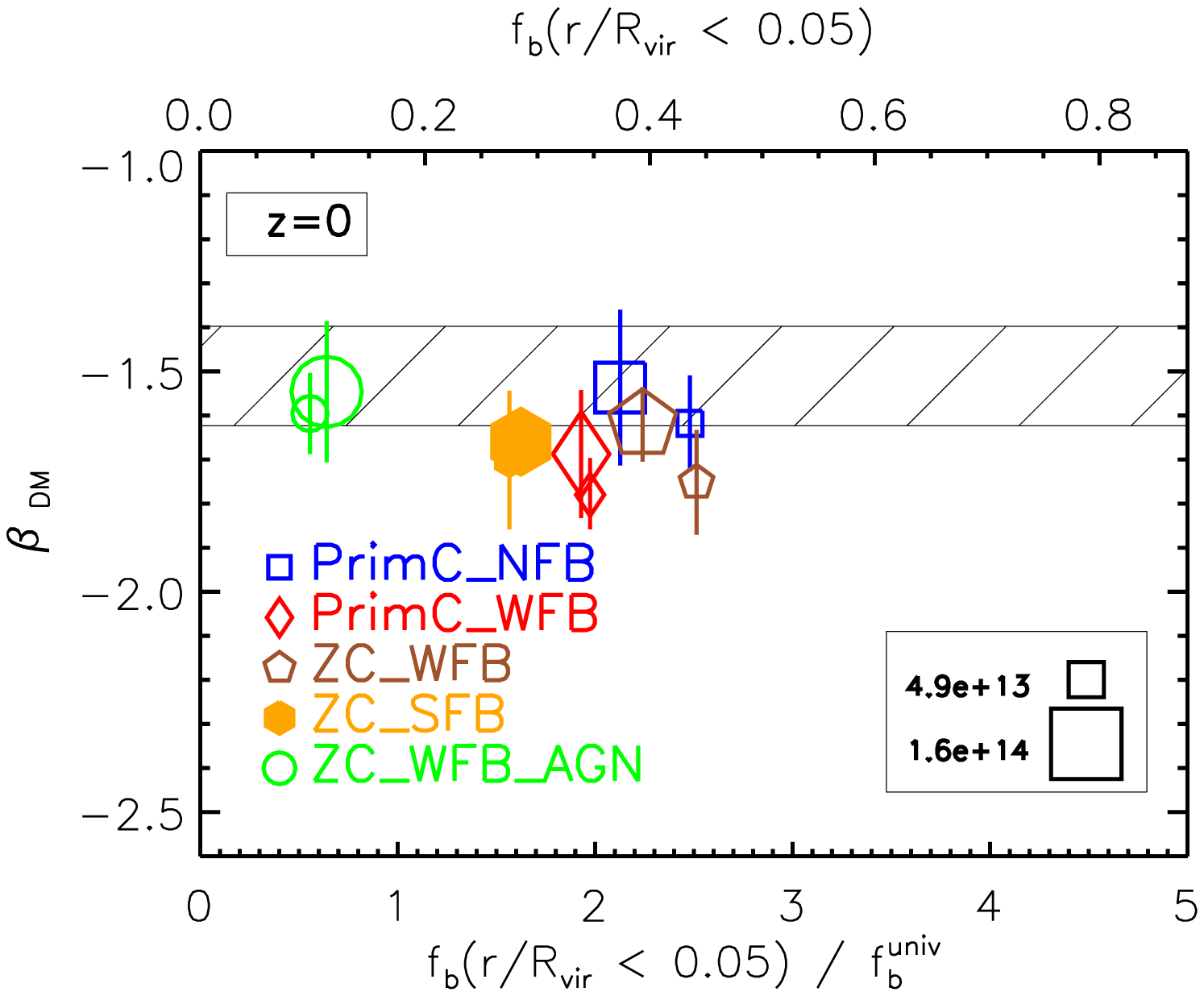, scale=0.45}
  \caption{The median inner ($0.025\leq r/R_{\rm vir} \leq 0.05$) power law slope of the DM density 
  profile as a function of the median baryon fraction within $r = 0.05\,R_{\rm vir}$, for different 
  simulations at $z=2$ (top) and $z=0$ (bottom).
  The symbols (and colours) represent different simulations while the
  symbol size indicates the mass 
  range. The vertical error bars illustrate the quartile scatter. The black hatched region represents 
  the quartile spread of the DMONLY simulation. Generically, higher
  central baryon concentrations yield steeper inner DM density profiles.}
  \label{fig:betafbprofile}
\end{figure}

It is clear that the main driver of the change in the (inner) DM profile is the 
condensation of gas
to smaller scales than the DM, thereby increasing both the
local baryon fraction and the DM concentration in the now-deepened
potential well.
If we assume that DM is pulled inwards by the condensing baryons, then we 
should expect some relation between the DM density profile and the baryon 
fraction within the inner region. We test this in Fig.~\ref{fig:betafbprofile} by plotting the median 
inner power law slope of the DM profile, 
$\beta_{\rm DM}$,  as a function of the median $f_{\rm b}(r/R_{\rm vir} \le 0.05)$ value. 
The quartile spread in $\beta$ values for the DMONLY run is also illustrated, as
the hatched region. We remind the reader that in the inner profile study we utilised a well-resolved
subsample of the full halo catalogue, such that $R_{\rm P03} \le 0.025 R_{\rm vir}$.

For the $z=0$ {\it Group} and {\it Cluster} haloes (bottom panel),
$f_{\rm b}$ varies by more than a factor of 3 in the inner region of the haloes and yet the
resultant inner slope stays within 20 per cent of the DMONLY value. 
There is a tentative steepening of the inner profile with 
increasing baryon fraction, but the steepest profile found in the 
simulation (PrimC\_WFB) does not have the largest central baryon fraction,
as already discussed in Section~\ref{sec:densityprofile_z0}.

Our $z=2$ {\it Galaxy} sample (top panel in Fig.~\ref{fig:betafbprofile}) does demonstrate
a clear and significant trend of steepening inner DM density profile with 
increasing central baryon fraction. The lowest baryon fractions are found in the 
ZC\_WFB\_AGN simulation for which the DM haloes are indistinguishable
from those in DMONLY. In models PrimC\_NFB and ZC\_WFB the 
baryon-dominated central regions generate nearly isothermal
($\beta=-2$) inner DM density
profiles.

\section{Halo Concentrations}
\label{sec:fitprofiles}

For an NFW halo of a given mass, the DM density profile is specified entirely by one parameter,
the concentration. In this section, we measure and compare NFW concentrations for the DM 
profiles of haloes in our simulations (Section~\ref{sec:cmrel}). 
We then consider the total mass profile of the halo by utilising simple, 
non-parametric measures of the concentration, based on ratios of density contrasts
at different scales (Section~\ref{sec:radratio}) and the velocity profile of the halo, in particular
the maximum velocity, in Section~\ref{sec:vc}. All haloes with a P03 convergence radius,
$R_{\rm P03}<0.05R_{\rm vir}$, are considered in this work on halo concentrations. 
This sets an effective mass limit of $M_{\rm vir}>8\times 10^{10}\hMsol$ at $z=2$ and 
$M_{\rm vir}>5\times 10^{12}\hMsol$ at $z=0$. The sample of haloes includes over 500 objects at 
$z=0$ and $\sim 20$ ($\sim 150$) systems at $z=2$ in
the simulation volumes with comoving length 100 (25) $h^{-1}\,{\rm Mpc}$.

\subsection{DM profile: NFW concentrations}
\label{sec:cmrel}

A well established result from $N$-body simulations~\citep{Bullock:01} is
that the NFW concentration of the DM halo is anti-correlated with its mass (for the latest results 
see~\citealt{Neto:07}; \citealt{Duffy:08b};~\citealt{Gao:08}
and~\citealt{Maccio:08}) and takes on a power-law form
\begin{equation}\label{eqn:powerlaw}
c_{\rm vir} = A_{\rm vir} (M_{\rm vir}/2\times 10^{12} h^{-1}\, {\rm M_{\odot}})^{B_{\rm vir}}\,,
\end{equation}
where $B_{\rm vir}$ is close to $-0.1$ when fit to data over nearly five orders of magnitude in 
mass. 
However, it is not clear how much this trend, which is primarily driven by the formation
time of the halo, is modified by the presence of baryons. 
Observations of X-ray luminous groups and 
clusters~\citep{Buote:07,SchmidtAllen:07} suggest a steeper 
dependence of concentration on mass than the DM only simulations predict, 
as was pointed out by \citet{Duffy:08b}. 
This is primarily due to observed groups having $\approx 30$ per cent higher concentrations than 
the simulated objects (the concentrations of clusters were in good agreement with
the simulations if a subsample of dynamically-relaxed haloes was used
in the comparison). It is therefore important to check whether the
inclusion of baryons can bring theory and observations into agreement.

\begin{figure}
  \epsfysize=2in \epsfxsize=4in
  \epsfig{figure=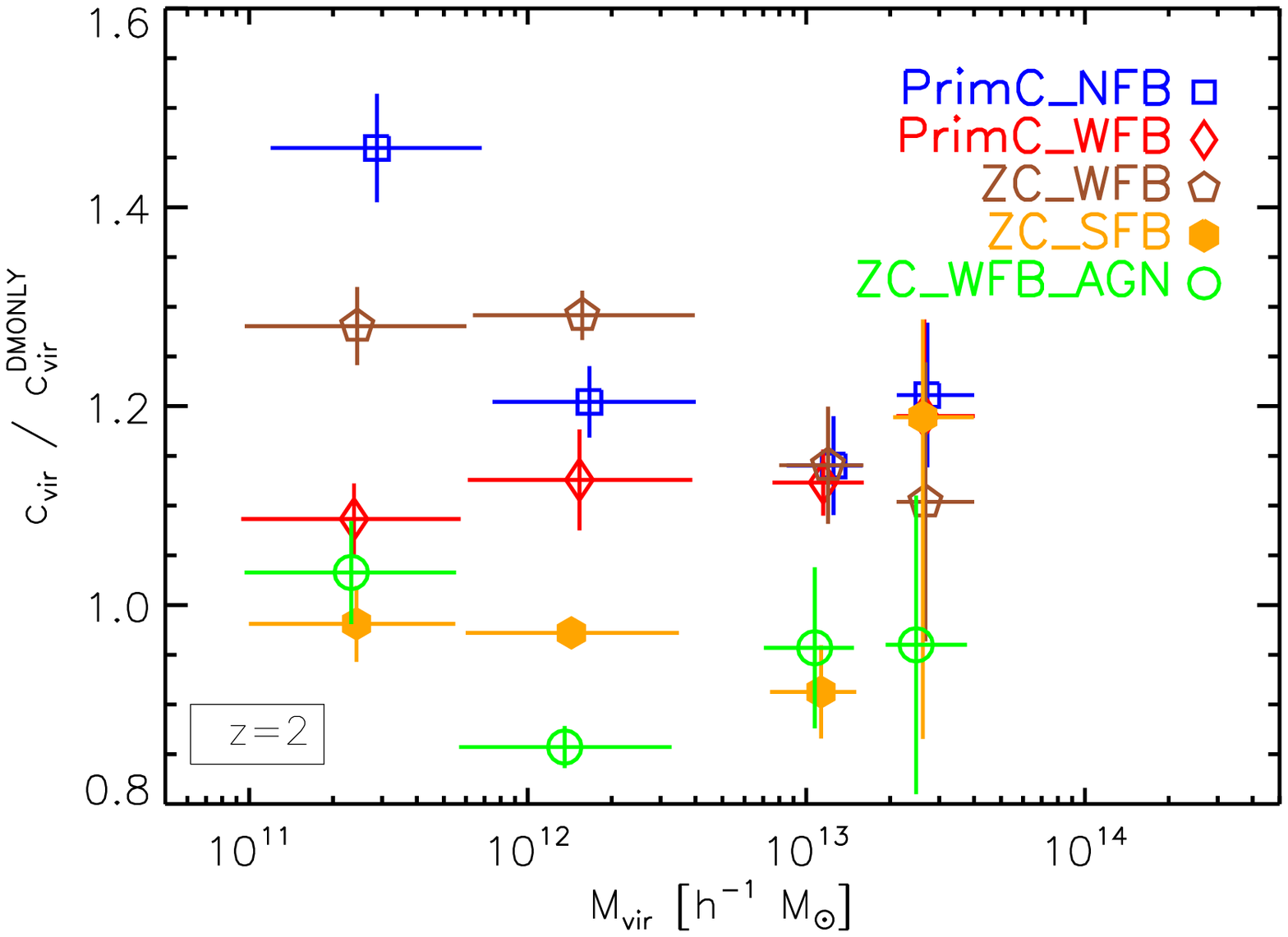, scale=0.45} 
  \epsfig{figure=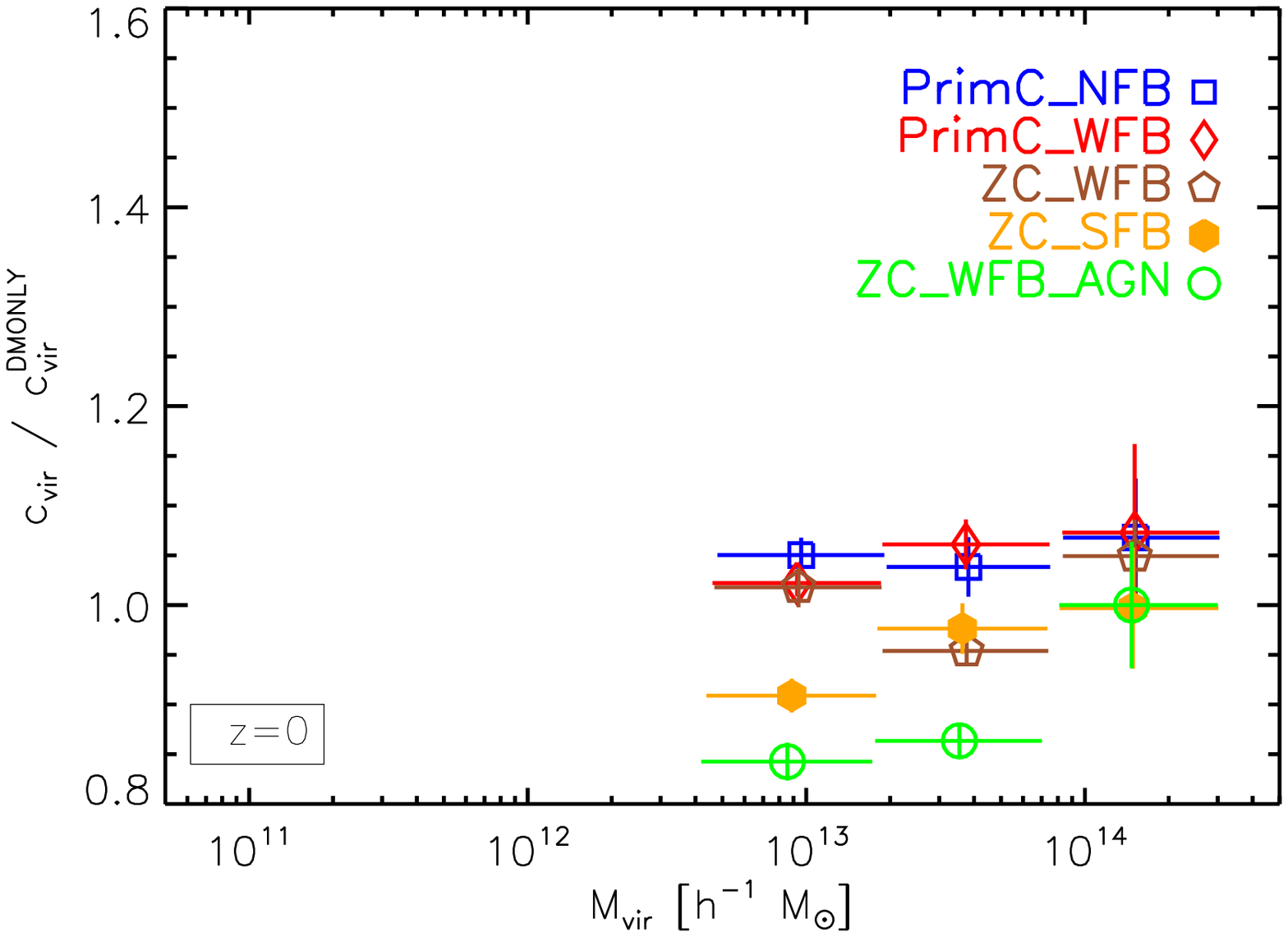, scale=0.45}
  \caption{We plot the NFW DM halo concentrations
    from the baryon simulations, normalised by the best fit
  concentration-mass relation from DMONLY, as a function of halo virial mass at $z=2$ (0) in the 
  top (bottom) panel. Values greater than unity indicate that the DM halo has 
contracted under the influence of the baryons.
 The points represent the 
   median concentration within each mass bin. The vertical error bars are the 68 per cent
   confidence intervals estimated by bootstrapping the haloes within each mass bin.
   The horizontal error bars indicate the mass range of each bin. Note
   that the gap in the mass coverage at 
   $z=2$ is where the two simulation volumes meet. In the absence of
   strong feedback and when metal-line cooling is included, baryons
   substantially increase the NFW DM 
   concentrations of {\it Galaxies}, but the effect on {\it Groups}
   and {\it Clusters} is much smaller. AGN and strong supernova feedback actually \emph{reduce} 
   the concentrations of {\it Groups}.}
  \label{fig:cvsmass}
\end{figure}

We will present concentrations relative to the
equivalent values for the DMONLY model. Note that our simulations
assume the WMAP3 cosmology, which has an 8 per cent
lower value of $\sigma_8$ than the WMAP5 value (0.74 versus 0.796)
assumed by \citet{Duffy:08b} and leads to somewhat smaller
concentrations at fixed mass\footnote{For reference, including all
  resolved haloes, our best-fit power law relations for DMONLY are
 $[A_{\rm vir}, B_{\rm vir}] = [7.8\pm{0.6}, -0.10\pm{0.03}]$ at $z=0$, and 
 $[A_{\rm vir}, B_{\rm vir}] = [3.7\pm{0.2}, 0.01\pm{0.03}]$ at $z=2$.}. As described in 
 Section~\ref{subsec:halo} we fit density profiles over the range $0.05 \leq r / R_{\rm vir} \leq 1$
 We first
 assess the goodness-of-fit of the NFW function when baryons are included. To do this,
 we compute the usual quantity 
 \begin{eqnarray}\label{eqn:sigfit}
\sigma_{\rm fit}^2  = \frac{1}{N_{\rm bins}} \sum_{i=1} ^{N_{\rm bins}} 
(\log_{10} \rho_{{\rm sim},i} - \log_{10} \rho_{{\rm NFW},i})^{2}\,,
\label{eqn:sigmafit}
\end{eqnarray}
where $N_{\rm bins}$ is the number of bins in our profile and $\rho_{\rm sim}$ and
$\rho_{\rm NFW}$ are the densities from the simulation
and the best-fit NFW profile respectively. For DMONLY we find that
$\sigma_{\rm fit} \approx 0.02$. For the runs with baryons the 
goodness-of-fit is similar (typically within 10 per cent) for {\it
  Group} and {\it Cluster} haloes. 
For smaller objects the difference is more pronounced for the simulations with 
high central baryon fractions, for which $\sigma_{\rm fit}$ increases by around a factor 2.

Shown in Fig.~\ref{fig:cvsmass} are the NFW concentrations of the DM haloes in
the runs with baryons, relative to the corresponding values from DMONLY, as
a function of halo virial mass. 
For the {\it Group} and {\it Cluster} haloes at $z=0$ (bottom row), the only simulation to show substantial ($>10$ per cent) 
deviations from the DMONLY values, is the simulation that includes AGN
feedback, ZC\_WFB\_AGN. For this model the NFW concentrations of
$M_{\rm vir} \sim 10^{13}\hMsol$ haloes are about 15 per cent lower.
For the run with strong stellar feedback, ZC\_SFB, the decrease is about 10 per cent.
In both cases the expulsion of gas has caused the
DM to expand relative to the DMONLY case (e.g.~\citealt{Hills:80}). 

The effect of baryons on the NFW concentrations of $z=2$ {\it Dwarf Galaxy}, {\it Galaxy} 
and {\it Group} haloes is shown in the top row of
Fig.~\ref{fig:cvsmass}. As before, the differences are more dramatic
for these lower mass, higher redshift objects. In the runs without strong feedback the
concentration increases, as expected. The increase is typically 10-20
per cent for {\it Groups}, but can be as large as 50 per
cent (when supernova feedback is absent) for {\it Dwarf Galaxies}. 
This dramatic increase is similar in magnitude to that found by~\citet{RomanoDiaz:09} for the concentration of a $M_{\rm vir} \approx 10^{12}\hMsol$ halo. 
As was the case at $z=0$, in runs with effective feedback the presence of
baryons makes little difference to the concentration of the DM; the maximum effect being a 
\emph{decrease} of $\sim 15$ per cent for {\it Galaxies} in ZC\_WFB\_AGN.

We checked how the results change when only {\it relaxed} haloes are selected (as defined
in \citealt{Duffy:08b}). As was found in previous work \citep[e.g.][]{Duffy:08b}, 
the average concentration of the haloes increases, but the power-law relation
between concentration and mass remains the same within the errors.

We have also checked explicitly that the concentration of a halo increases with its central
baryon fraction. This is indeed the case for all simulations and mass scales, except
for the {\it Group} and {\it Cluster} haloes at $z=0$, in the PrimC\_WFB model. As discussed in
the previous section, this run has haloes on these mass scales with
anomalously high central DM densities (as compared with the run with primordial cooling and no feedback, PrimC\_NFB).

In~\citet{Duffy:08b} we demonstrated that the inferred NFW concentrations of groups observed in 
X-rays at $z=0$ are $~ 30$ per cent more concentrated than predicted from DM only simulations.
It was suggested that the inclusion of baryons would alleviate this discrepancy through the 
contraction of the DM halo. As is clear from Fig.~\ref{fig:cvsmass}, however, the largest increase of 
any physics scheme is still less than $10$ per cent. Moreover, for those schemes which reproduce 
the observed $z=0$ stellar fractions (shown in Fig.~\ref{fig:fstar_m500}), the strength of feedback 
is such that 
we actually \emph{reduce} the concentration of the {\it Groups} relative to the DMONLY simulation.
As an additional issue, the trend of the concentration ratio with mass is positive in the strong
feedback schemes which may further increase the disagreement with observations.
Clearly, the inclusion of baryons does not resolve the problem.
It is thus important to check whether observational biases or selection effects can account for the
mismatch between theory and observation.

\subsection{Total mass profile: Non-parametric concentrations}
\label{sec:radratio}

\begin{figure}
  \epsfysize=2in \epsfxsize=4in
  \epsfig{figure=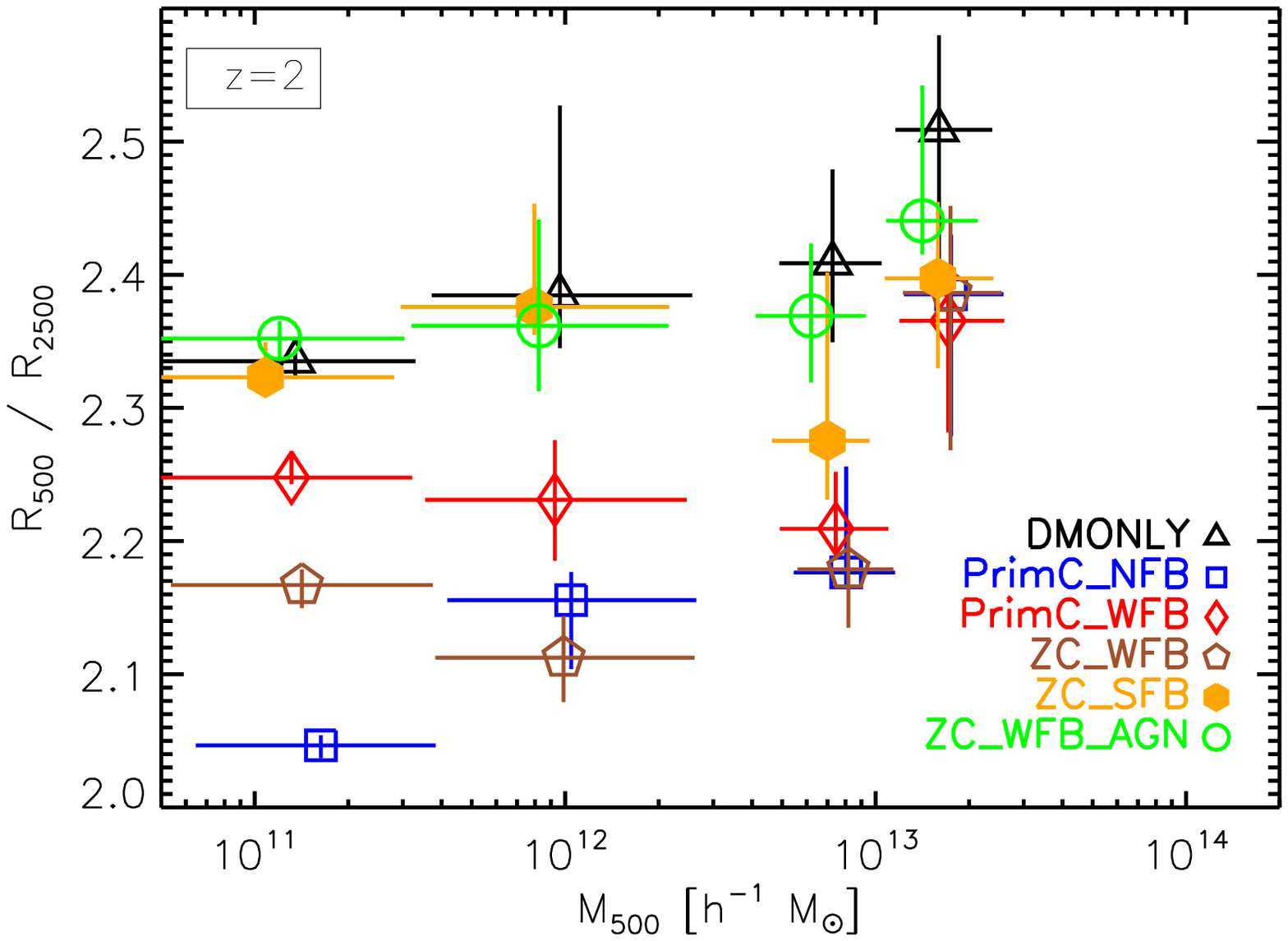, scale=0.4}
  \epsfig{figure=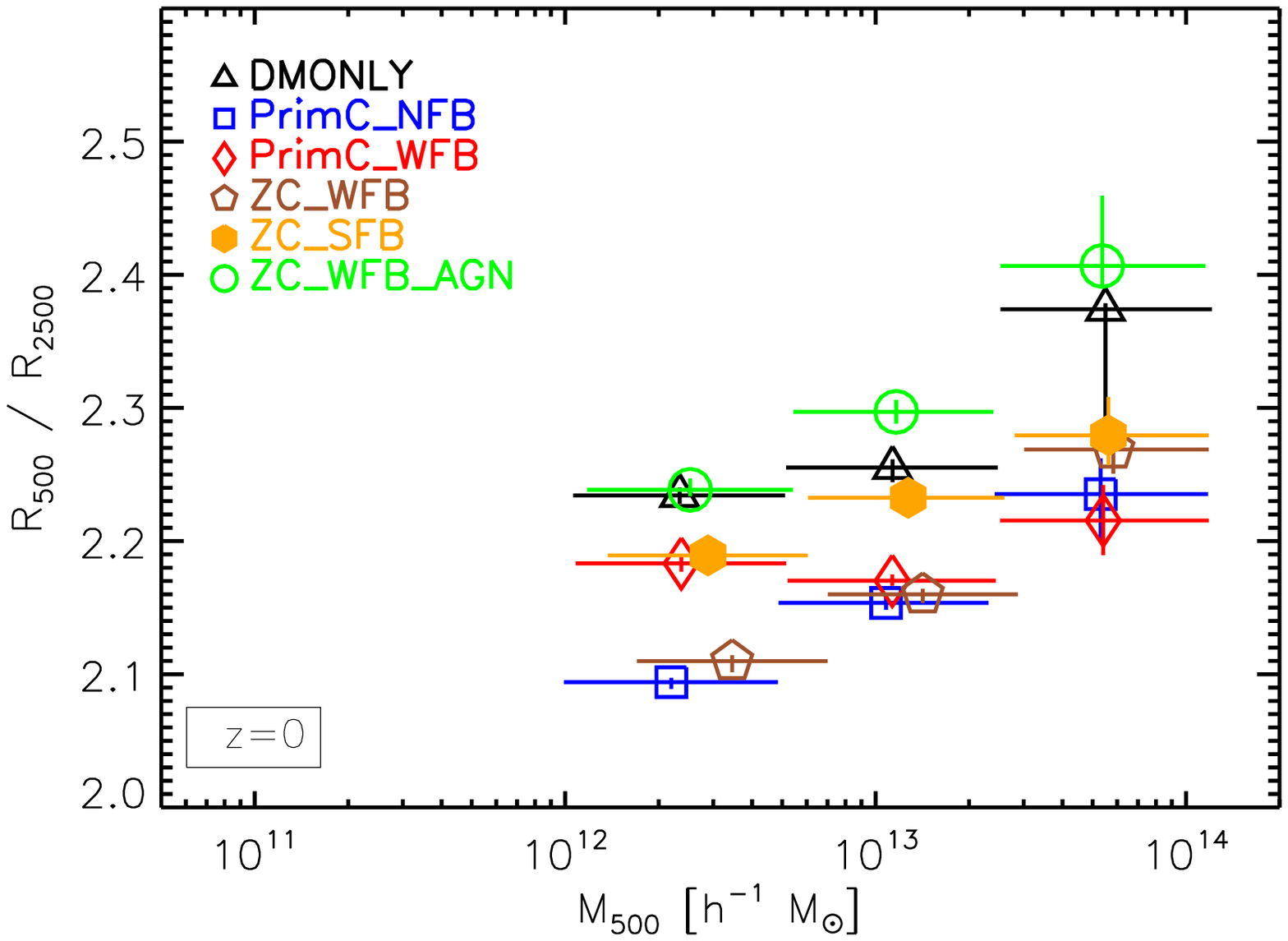, scale=0.4}
  \caption{Here we plot the median ratio of
  the spherical overdensity radii $R_{500}$ and $R_{2500}$ as a 
  function of halo mass, $M_{500}$ in this case, at $z=2$ (top) and $0$ (bottom).
  This ratio is a useful non-parametric measure of the concentration of a halo. The vertical 
  error bars are 68\% confidence limits about the medians, estimated by bootstrapping 
  the samples within each bin, and the horizontal values represent the mass range of the bin. Note 
  that higher values of the ratio indicate \emph{lower} concentrations. All 
  simulations show a positive trend with mass; this is indicative of the decreasing amount
  of baryons able to cool and condense as the virial temperature of the halo increases. The offset
  in the normalisation between simulation schemes is larger than the scatter, allowing the
  use of this ratio as a nonparametric concentration proxy.}
  \label{fig:r500r2500}
\end{figure}

Although the NFW profile is a reasonable approximation to the DM profile at large radii 
($r/R_{\rm vir} > 0.05$, as shown in Fig.~\ref{fig:allspeciesdensplot}), the measurement
of its concentration parameter requires the shape of the profile to
be constrained over a range of scales, with accurate removal of the baryonic mass profile.
A simpler, and thus more easily achievable method from an observational point
of view, is to consider the entire halo, DM and baryon components together, and measure the 
mass/radius ratio of a halo at two different spatial scales. 

In Fig. \ref{fig:r500r2500} we plot the ratio of two radii that are commonly used by observers (e.g.
in observations of X-ray groups and clusters), $R_{500}/R_{2500}$\footnotemark, as a function of  halo mass. 
\footnotetext{Typical values for these radii in terms of $R_{\rm vir}$ are $\approx 0.5 \,(0.6)$ and 
$0.2 \,(0.3)$ for $R_{500}$ and $R_{2500}$ at $z=0 \,(2)$ respectively. The values will change by 10 
per cent dependent on the simulation physics due to the baryonic back-reaction.}
All runs demonstrate a positive, albeit weak, dependence on mass with significant run-to-run 
variations in the normalisation. (Note that runs with higher central baryon fractions will typically 
have {\it lower} $R_{500}/R_{2500}$ ratios, because the value of $R_{2500}$ grows as a result of 
the increased central mass.) 
The deviation from DMONLY is at the sub-25 per cent level and much smaller if the feedback is 
strong.
The differences between the models are qualitatively similar to those for the NFW concentrations 
discussed in the previous section. Note, however, that contrary to the NFW concentrations, the
non-parametric total matter concentrations are never significantly reduced relative to the DMONLY
case.

\subsection{Concentration measures: circular velocity}
\label{sec:vc}

A key, and relatively easily observed measure for the structure of a halo is the circular velocity,
$v_{\rm c}(r) = (GM(< r)/r)^{1/2}$, in particular its maximal value, $v_{\rm max}$. 
Like $R_{500}/R_{2500}$, it can be more robustly determined than the NFW concentration. 
The creation of realistic velocity profiles for galaxies has, however, been a long standing issue in 
simulations within the CDM paradigm. 
For $N$-body only simulations, the maximum 
velocity is well approximated by the analytic solution to the NFW profile 
$v_{\rm max} = v_{\rm c}(r \approx 2.17 r_{\rm s})$~\citep{ColeLacey96, NFW:96}, with the final
result that $v_{\rm max} \approx V_{\rm vir} = (G M_{\rm vir}/ R_{\rm vir})^{1/2}$. 
Observations find that the maximum velocity in the disk is similar to the virial 
velocity~\citep[e.g.][]{Dutton:05}.
Typically, however, when simulations include baryons, either explicitly or through adiabatic 
contraction models, the haloes will have a maximum velocity that is a factor of two higher than 
the halo virial velocity~\citep[e.g.][]{Navarro:00, Dutton:07, Pedrosa:10}. This increase in the
velocity ratio is a consequence of the contraction of the halo in the presence of a significant
central baryon component, with a large velocity ratio indicative of a high NFW concentration.

Recently,~\citet{Pedrosa:10} investigated the circular velocity profile in a resimulation of a single
high resolution galaxy for various physics implementations. 
They found that there was a positive correlation between the 
ratio $v_{\rm max}/V_{\rm 200}$ and the `sharpness' of the DM density profile\footnotemark, 
indicative of the contraction of the halo in the presence of baryons. Furthermore, effective 
feedback was necessary to obtain a realistic, i.e. low, velocity ratio in their galaxy.
In addition to {\it galaxies} we have investigated a larger mass range, {\it groups} and 
{\it clusters}. We also lend statistical weight to conclusions concerning the importance
of the various physics implementations by examining a large sample of haloes.
A more extensive investigation of the rotation properties of the OWLS galaxies at $z=2$, in
addition to $v_{\rm max}$, is being performed in a separate study (Sales et al., in prep).
\footnotetext{The sharpness of the DM halo is characterised by the rate at which the density
profile becomes shallower towards the halo centre. This is explicitly modelled in the Einasto profile~\citep{Einasto:65} as an additional parameter to the NFW in the form of a rolling power law.}

\begin{figure}
  \begin{center}
    \epsfig{figure=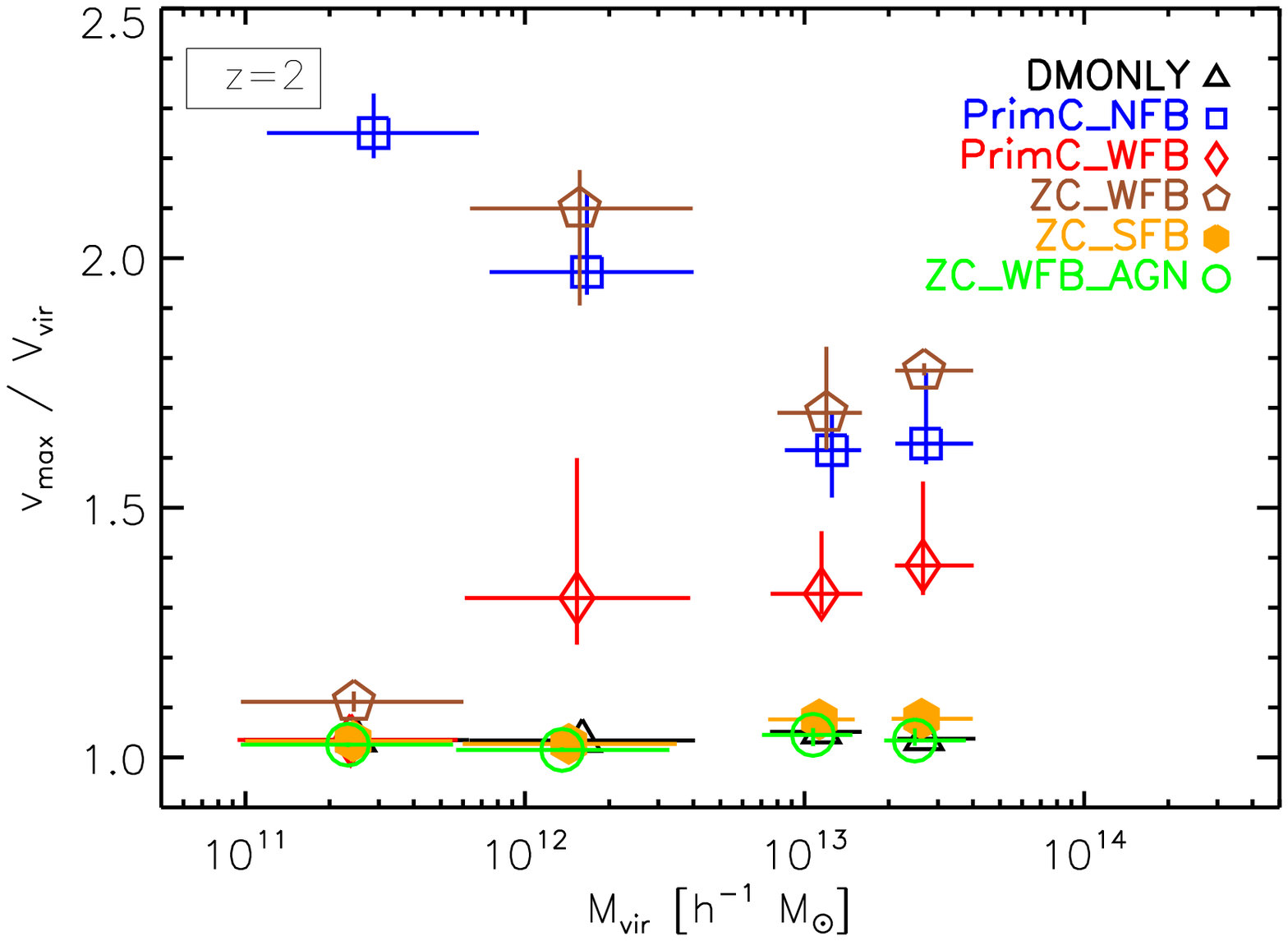, scale=0.45}
    \epsfig{figure=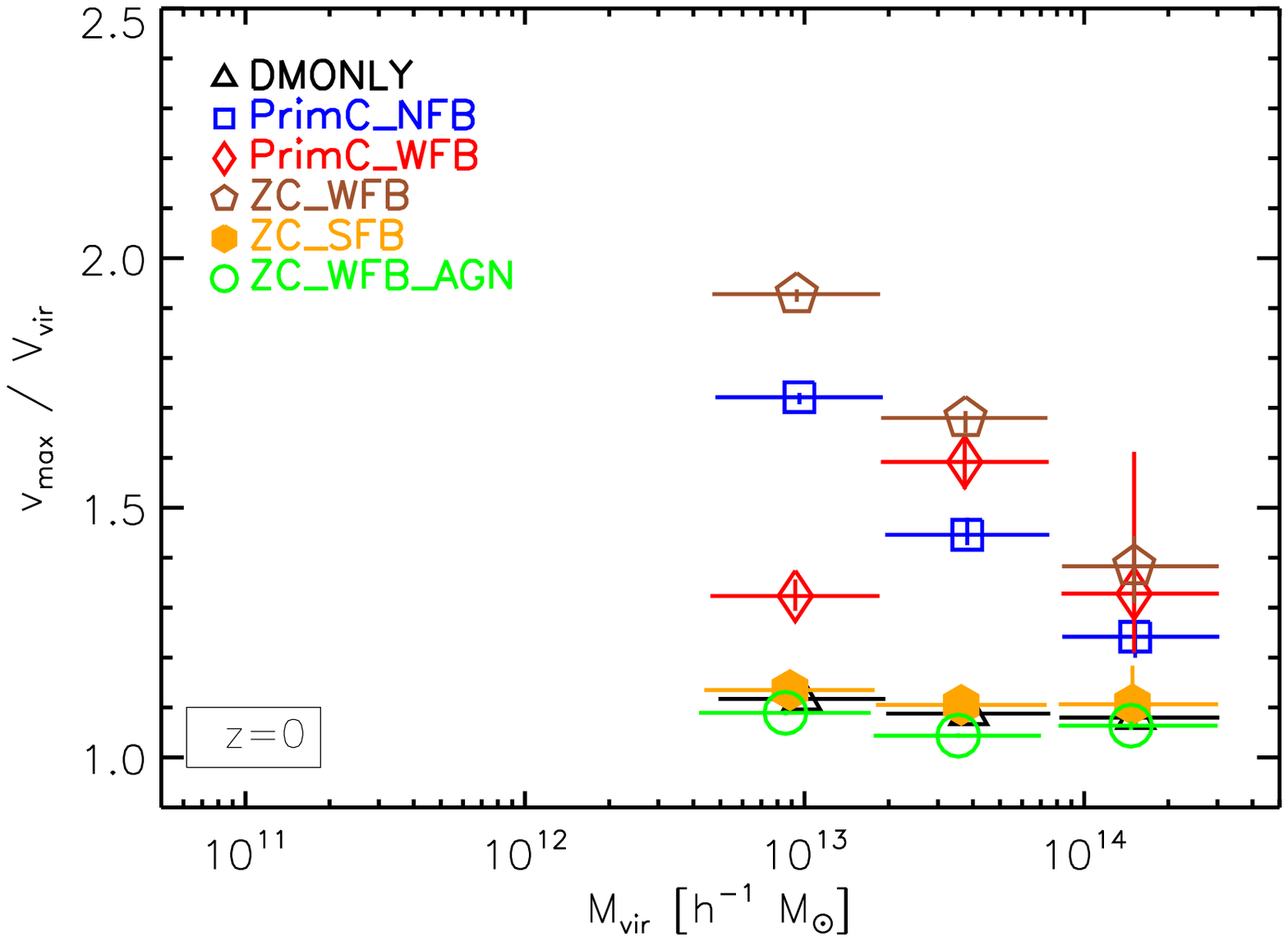, scale=0.45}
    \caption{We plot the maximum circular velocity of the halo, relative to the circular velocity at the 
    virial radius, $R_{\rm vir}$, as a function of virial mass at $z=2$ (0) in the top (bottom) panel. 
    This measure is sensitive to the degree of contraction of the halo in the presence of baryons.
    Each point 
    corresponds to the median value for haloes within a given mass bin. Vertical error bars 
    are bootstrap estimates of the 68\% confidence interval for each bin and the horizontal bars
    are the bin widths.
    A number of familiar halo responses is easily seen in this figure, with {\it galaxies} more 
    sensitive to the presence of the baryons than higher mass systems.
    The very high maximum velocity in PrimC\_NFB indicates the importance of feedback
    in forming observed galaxies, whose maximum circular velocity in the disk is
    similar to the virial velocity~\citep{Dutton:05}. 
    Furthermore, the sharp divergence between PrimC\_NFB and PrimC\_WFB at a critical mass
    of $10^{13} \hMsol$ indicates the mass range at which the weak feedback model becomes
     ineffective. Finally, the metal-line cooling has the overall effect of counteracting the 
      supernova feedback in {\it Galaxies} and {\it Groups} at $z=2$, seen by the agreement 
       between PrimC\_NFB and ZC\_WFB. The anti-correlation seen at $z=0$, for the simulations
        PrimC\_NFB and ZC\_WFB, is indicative of the rising halo virial temperature restricting the 
	 efficiency of gas cooling.
	 The strong feedback schemes are in close agreement to the DMONLY simulation 
	 predictions at all masses and redshifts, and are therefore necessary to recover observed
	 values.}
    \label{fig:vc}
  \end{center}
\end{figure}

In Fig. \ref{fig:vc} we show the maximum circular velocity, in units of the virial velocity, as a function 
of halo mass at $z=2$ ($0$) in the top (bottom) panel for various implementations of the 
cooling and feedback prescriptions. 
The most striking result is the good agreement between the
strong feedback schemes (and DMONLY) with $v_{\rm max} \approx V_{\rm vir}$ at all masses
and redshifts, and the significant offset for the other physics implementations. 
This divergence is reduced with increasing mass at all redshifts due to the strong anti-correlation of 
the velocity ratio with halo mass in the weak/no feedback runs. This is indicative of the reduction in 
gas cooling efficiency as the halo mass, and hence the virial temperature, increases.

At $z=2$ (top panel) there is a dramatic divergence in the velocity profile below $10^{13} \hMsol$
between the runs with primordial cooling, due to the supernova feedback.
When metal-line cooling is included, ZC\_WFB, the divergence 
with the no feedback model is reduced; the overall effect of the enhanced cooling is therefore to 
counteract the gas removal efforts of the supernovae.

At $z=0$ (bottom panel of Fig. \ref{fig:vc}) we once again find the counterintuitive result that the 
model with weak feedback, PrimC\_WFB, is more concentrated in {\it clusters} (but not in {\it 
groups}) than the model without feedback PrimC\_NFB. We also see the familiar effect of the 
metal-line 
cooling overcoming the weak supernova feedback when comparing PrimC\_NFB and ZC\_WFB 
models. 

It is clear that strong feedback is necessary if one wishes to limit the effect of the baryons in 
increasing the maximum circular velocity of the halo above the virial velocity.

\section{Adiabatic Contraction}
\label{sec:AC}

Having demonstrated that baryons can significantly influence the
DM density profile, we will now assess the degree to which this 
modification can be modelled as a secular adiabatic contraction of the
DM halo. We will test the models of 
\citet{Blumenthal:86}, henceforth B86, and G04, and will make use of
the publicly-available code {\sc contra} (G04).  

The B86 model for adiabatic contraction assumes that the DM halo 
is spherically-symmetric and that the particles are on circular
orbits so that we can think of the halo as a series of shells that can
contract but do not cross. Assuming that the baryons initially
trace the DM halo density 
profile and then fall slowly (i.e., such that the mass internal to
radius $r$ changes slowly compared to the orbital period at $r$)
towards the centre as they cool, we can compute 
the response of the dark matter shells. In
that case the dark matter particles conserve their angular momentum
and hence $r v_{\rm 
  c} \propto [r M(r)]^{1/2}$ is conserved, where $M(r)$ is the total
mass internal to $r$. We thus have 
\begin{eqnarray}
{M_{\rm dm,i}(r_{\rm i})r_{\rm i} \over 1-f_{\rm b}^{\rm univ}}  &=& 
\left [M_{\rm dm,f}(r_{\rm f}) + M_{\rm b,f}(r_{\rm f})\right ] r_{\rm f},
\\
&=& \left [M_{\rm dm,i}(r_{\rm i}) + M_{\rm b,f}(r_{\rm f})\right ] r_{\rm f},
\end{eqnarray}
where $M_{\rm dm,i}(r_i)$ is the initial DM profile, $M_{\rm
  b,f}(r_{\rm f})$ is the final baryon profile, and $r_{\rm
  f}$ is the final radius of the DM shell that was initially 
(i.e.\ before the baryons contracted) at $r_{\rm i}$. Note that we
made use of the equality $M_{\rm dm,f}(r_{\rm f}) = M_{\rm
    dm,i}(r_{\rm i})$, which holds because DM shells do not
cross. Assuming that we know both $M_{\rm b,f}(r_{\rm f})$ and 
$M_{\rm dm,i}(r_{\rm i})$, we can solve for $r_{\rm i}$ (and thus $M_{{\rm
    dm},i}(r_{\rm i}) = M_{\rm dm}(r_{\rm f})$) as a function of
$r_{\rm f}$. In the following we will make the standard assumption
that $M_{\rm dm,i}(r_{\rm i})$ is given by the density profile
(reduced by $(1-f_{\rm b}^{\rm univ})$) of the
corresponding DMONLY halo.

\begin{figure*}
  \epsfysize=2in \epsfxsize=4in
  \begin{tabular}{ccc}
  \epsfig{figure=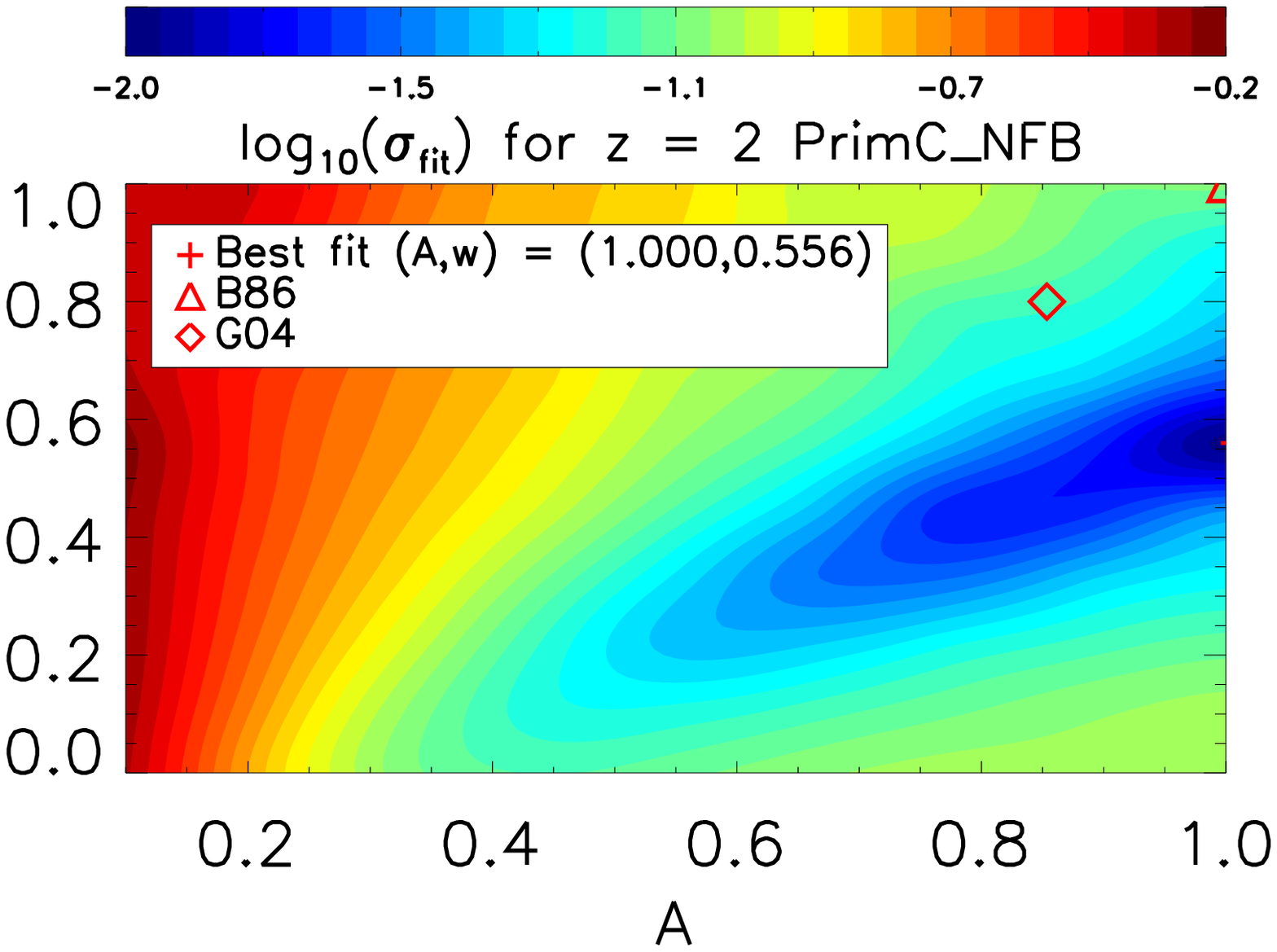, scale=0.32} & 
  \epsfig{figure=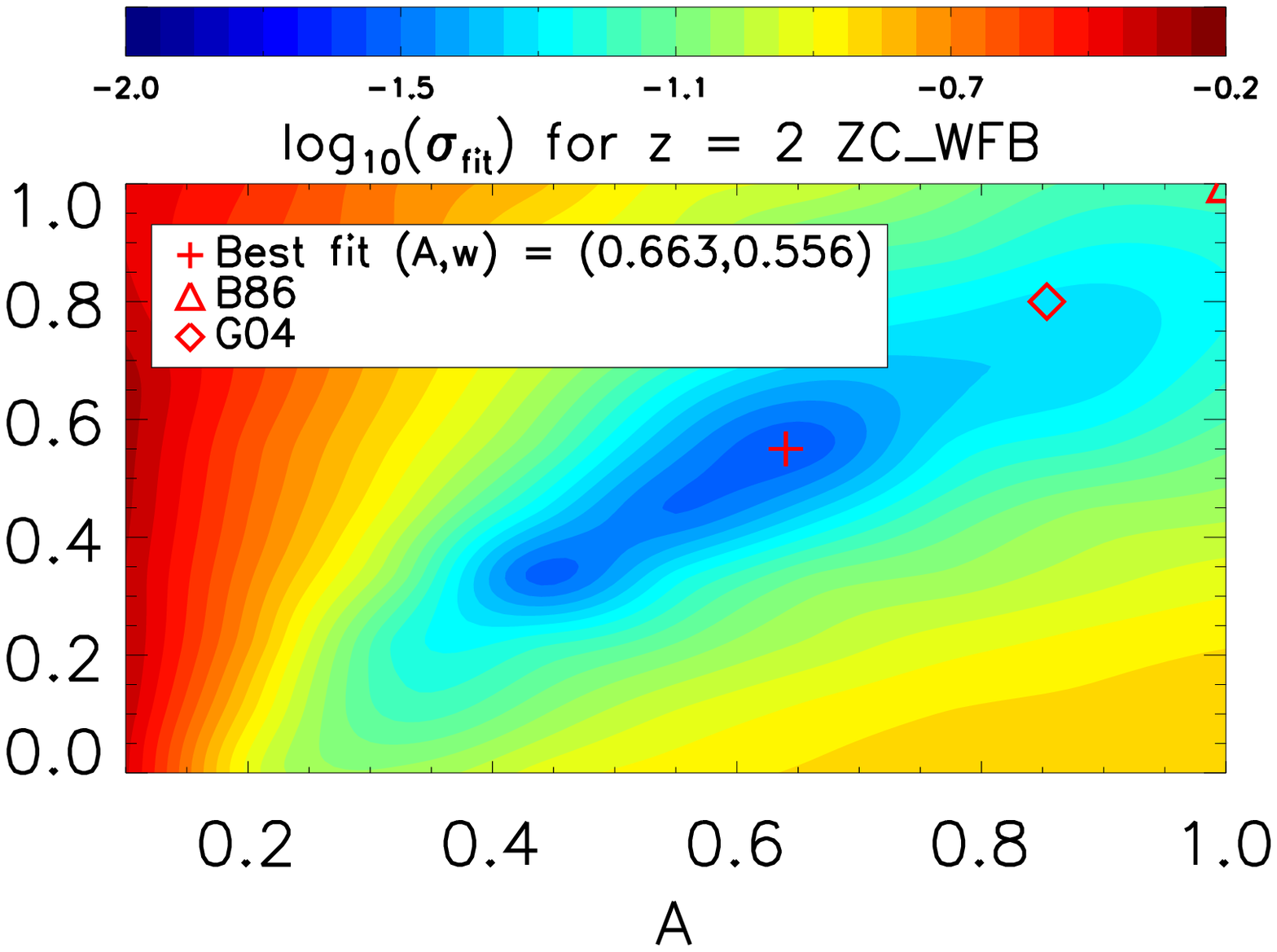, scale=0.32} & 
  \epsfig{figure=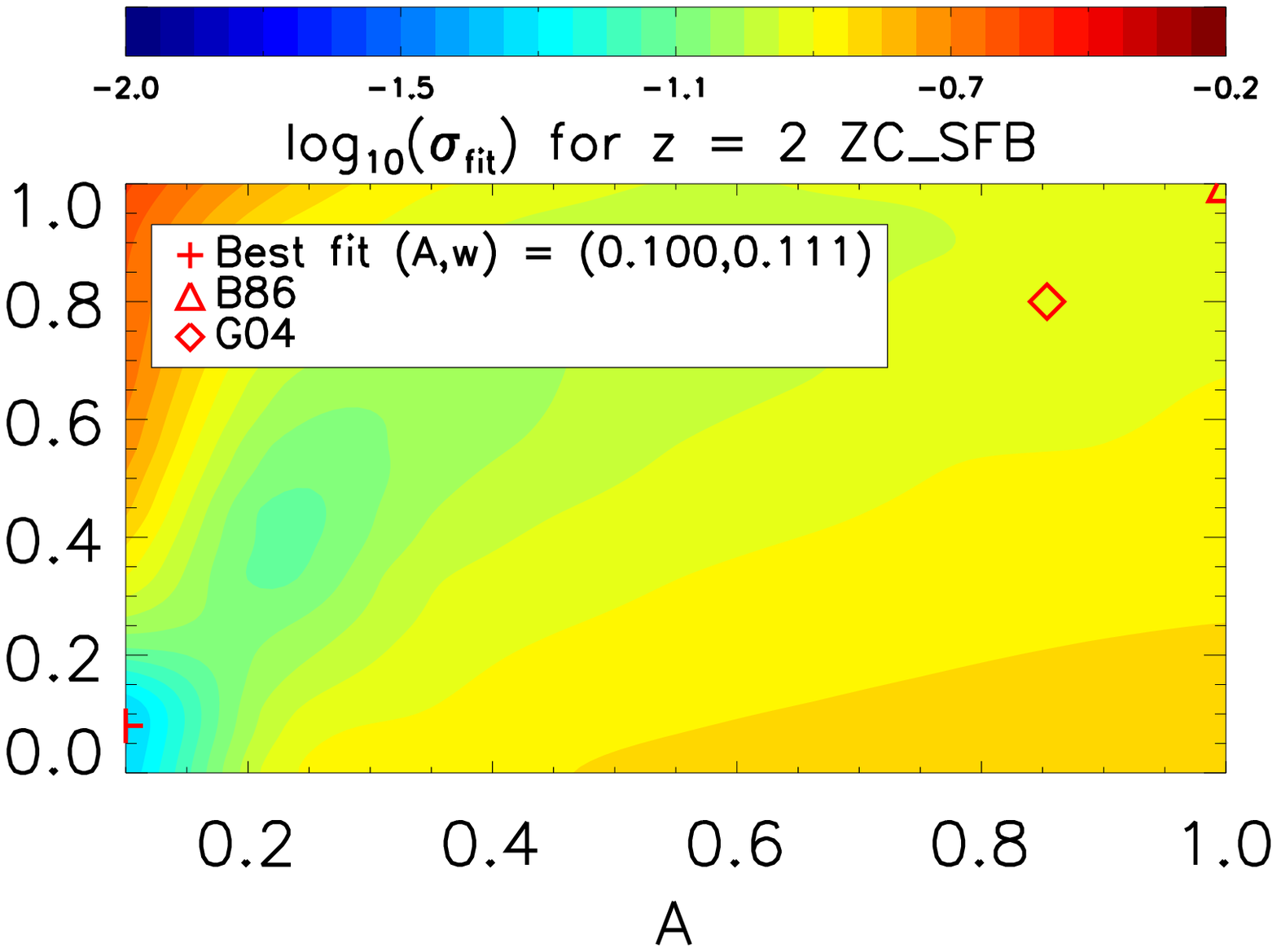, scale=0.32} \\  
  \epsfig{figure=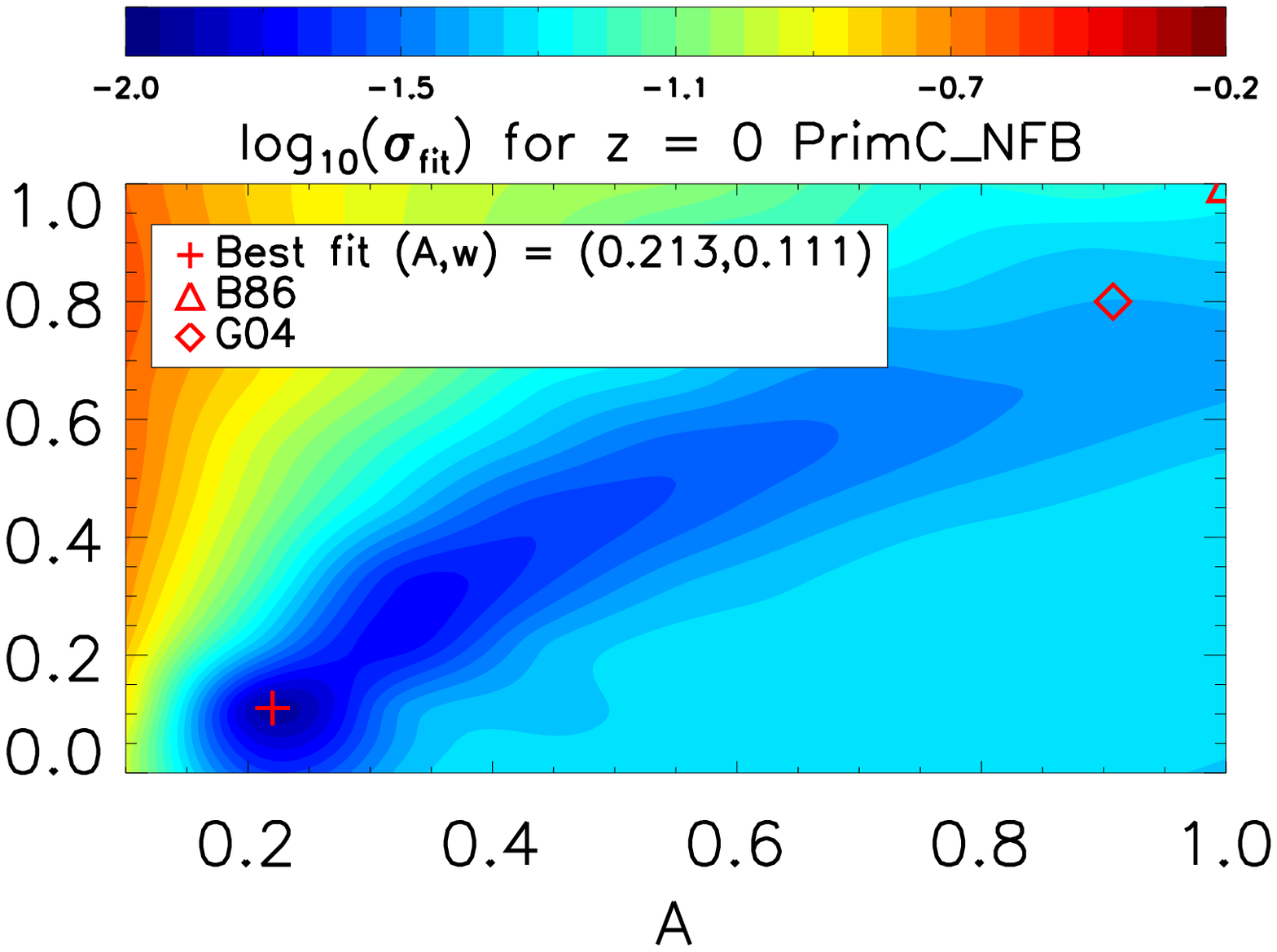, scale=0.32} &
  \epsfig{figure=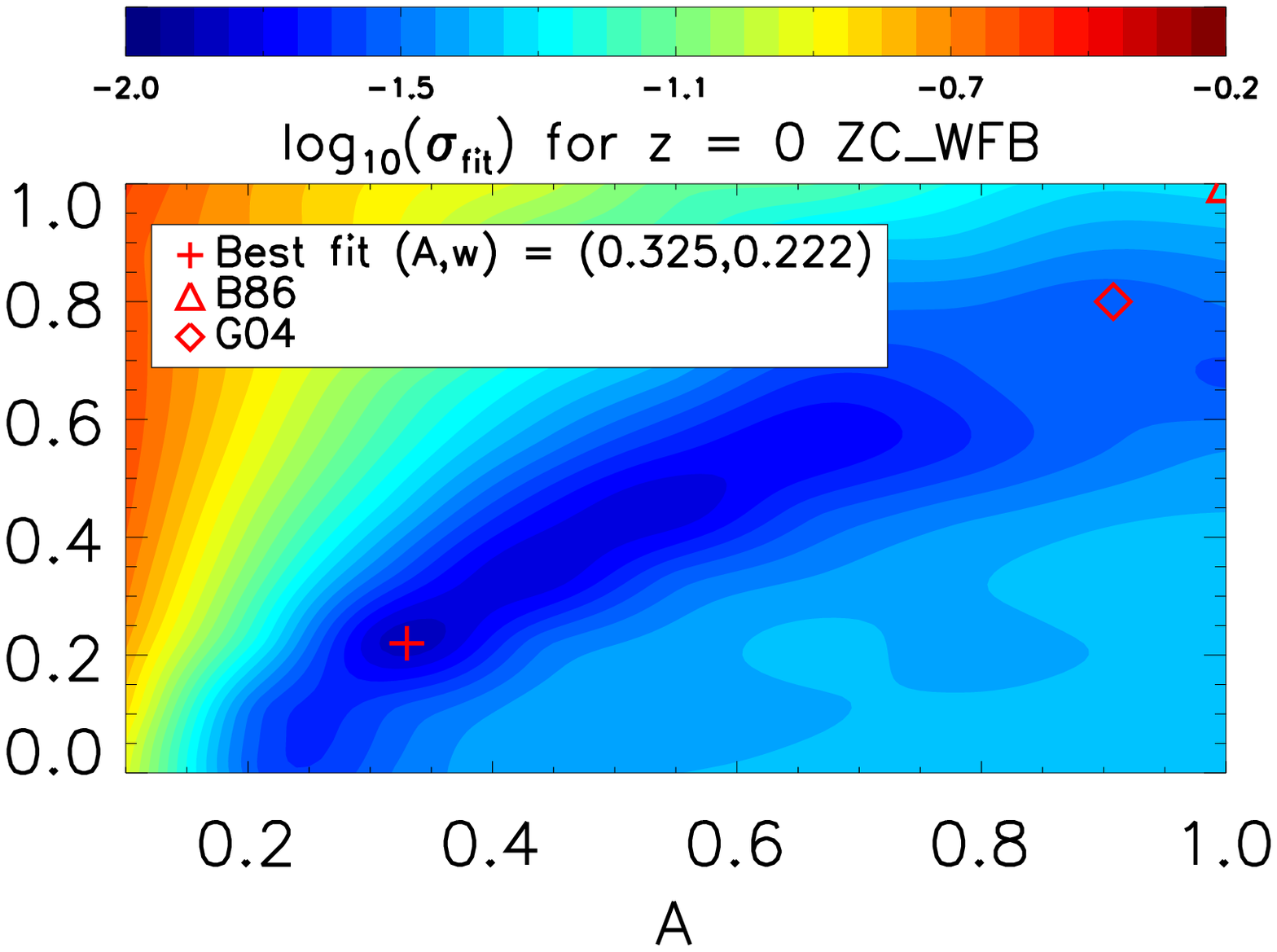, scale=0.32} &
  \epsfig{figure=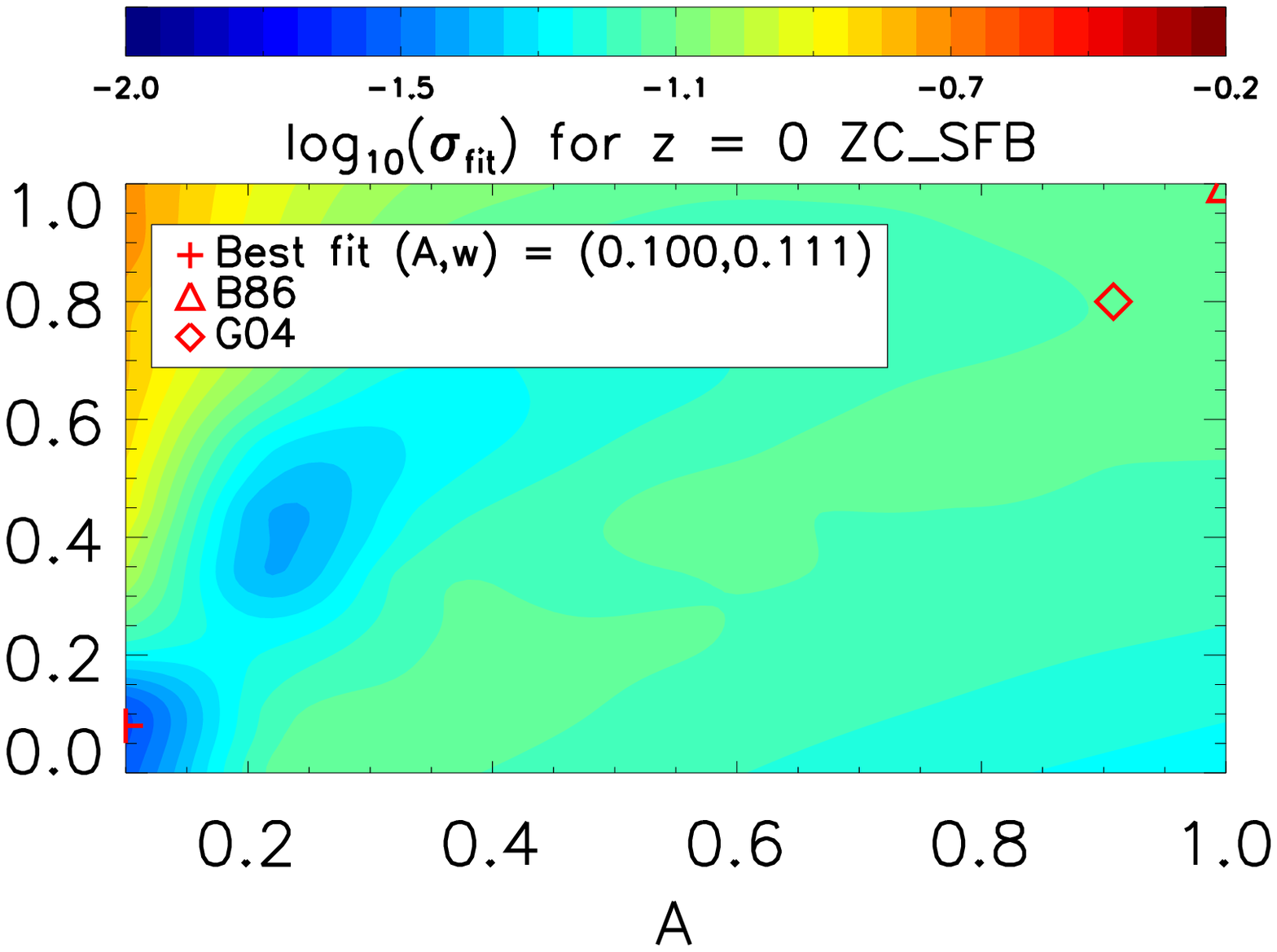, scale=0.32} \\
  \end{tabular}
  \caption{The goodness of fit,  $\sigma_{\rm fit}$, of the G04 model
    for adiabatic contraction as a function of the parameters $A$ and
    $w$, in logarithmic units. The best-fit parameter combination is indicated by the red cross. 
    The top and bottom rows show results at $z=2$ and $z=0$
    respectively. From left-to-right we show the 
    results for the PrimC\_NFB, ZC\_WFB and ZC\_SFB simulations respectively. 
    For each simulation the results are averaged over haloes matched
    to the DMONLY 
    haloes with virial masses in the range $5 - 50 \times 10^{11}$ 
    ($3 - 60 \times 10^{13}$) $\,h^{-1}\,{\rm M_{\odot}}$ for the $z=2$ ($z=0$) 
    samples.  
    The parameter values corresponding to the models of B86 ($A=w=1$,
    hence top right corner) and G04 are shown as asterisk and open
    diamonds, respectively. The two parameters are substantially
    degenerate. The 
    best-fit values depend on halo mass, redshift, and on the
    implemented baryonic physics.}
  \label{fig:avsw}
\end{figure*}

G04 found that this method did not provide
a satisfactory description of their numerical simulations. Because the
DM particles are typically not on circular orbits, they
suggested replacing $M(r)r$ by $M(\bar{r})r$, where $\bar{r}$ is the
{\it orbit-averaged} radius, which they found can be approximated as
\begin{equation} 
\bar{r}=R_{\rm vir} A(r/R_{\rm vir})^w,
\end{equation}
where $A=0.85$ and $w=0.8$ (c.f.\ the B86 model which has $A=w=1$).
\citet{Gustafsson:06} extended this work and showed that the values of
$A$ and $w$ that best describe the difference between simulations with
and without baryons depend on halo mass and the baryonic physics
implementation. Furthermore, they showed that the best fitting values
of $A$ and $w$ generally differ substantially from the values that
provide a good fit to the actual mean radius $\bar{r}$ of the DM particle
orbits. This suggests that while the introduction of the two free
parameters $A$ and $w$ enables better fits, the underlying model does
not capture the relevant physics.
Here, we extend the work by \citet{Gustafsson:06} to a larger range of
mass scales and gas physics models.

\subsection{Best-fit G04 parameters}

\begin{figure*}
  \epsfysize=2in \epsfxsize=4in
  \begin{tabular}{ccc}
  \epsfig{figure=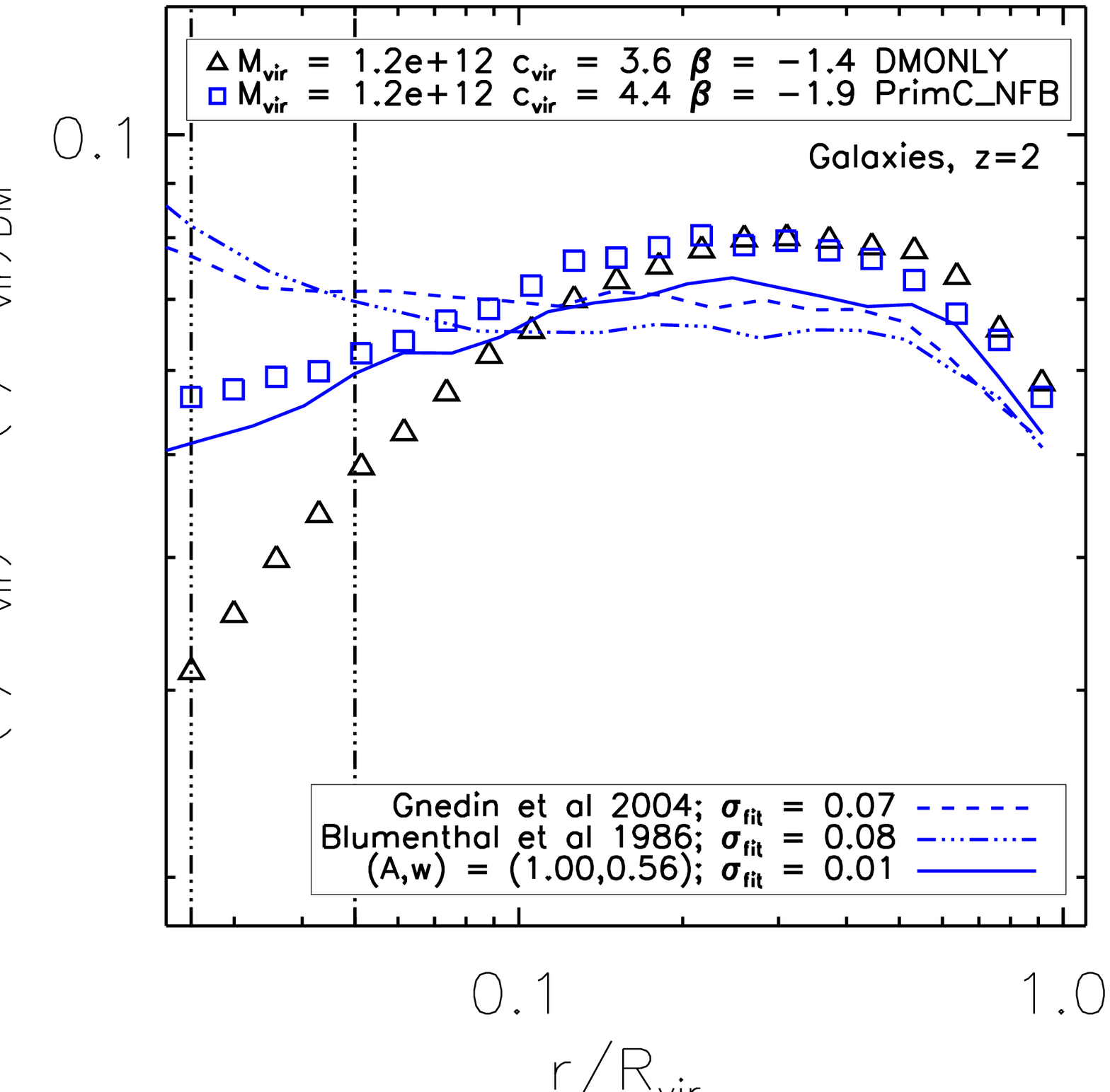, scale=0.3} &
  \epsfig{figure=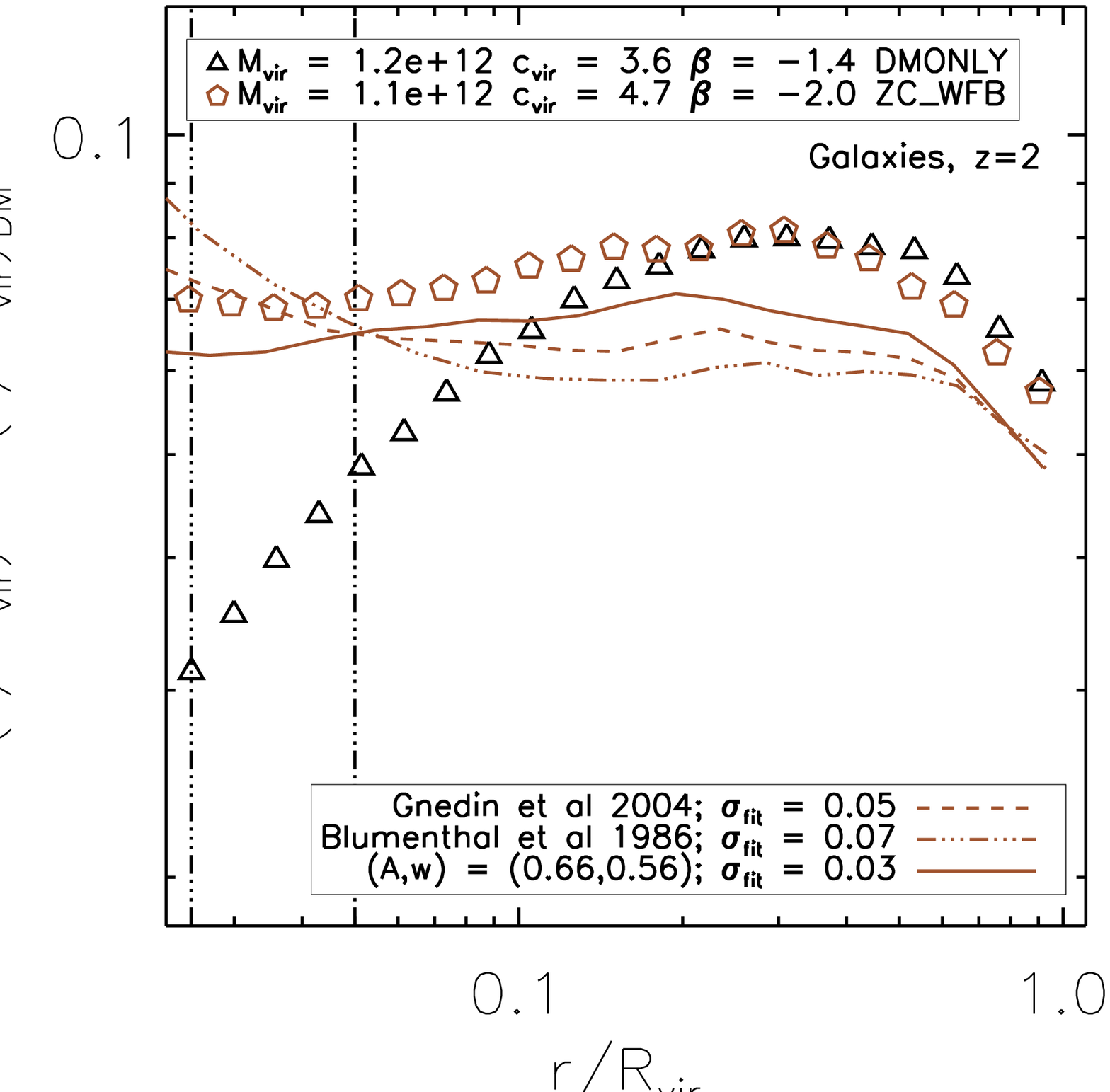, scale=0.3} &
  \epsfig{figure=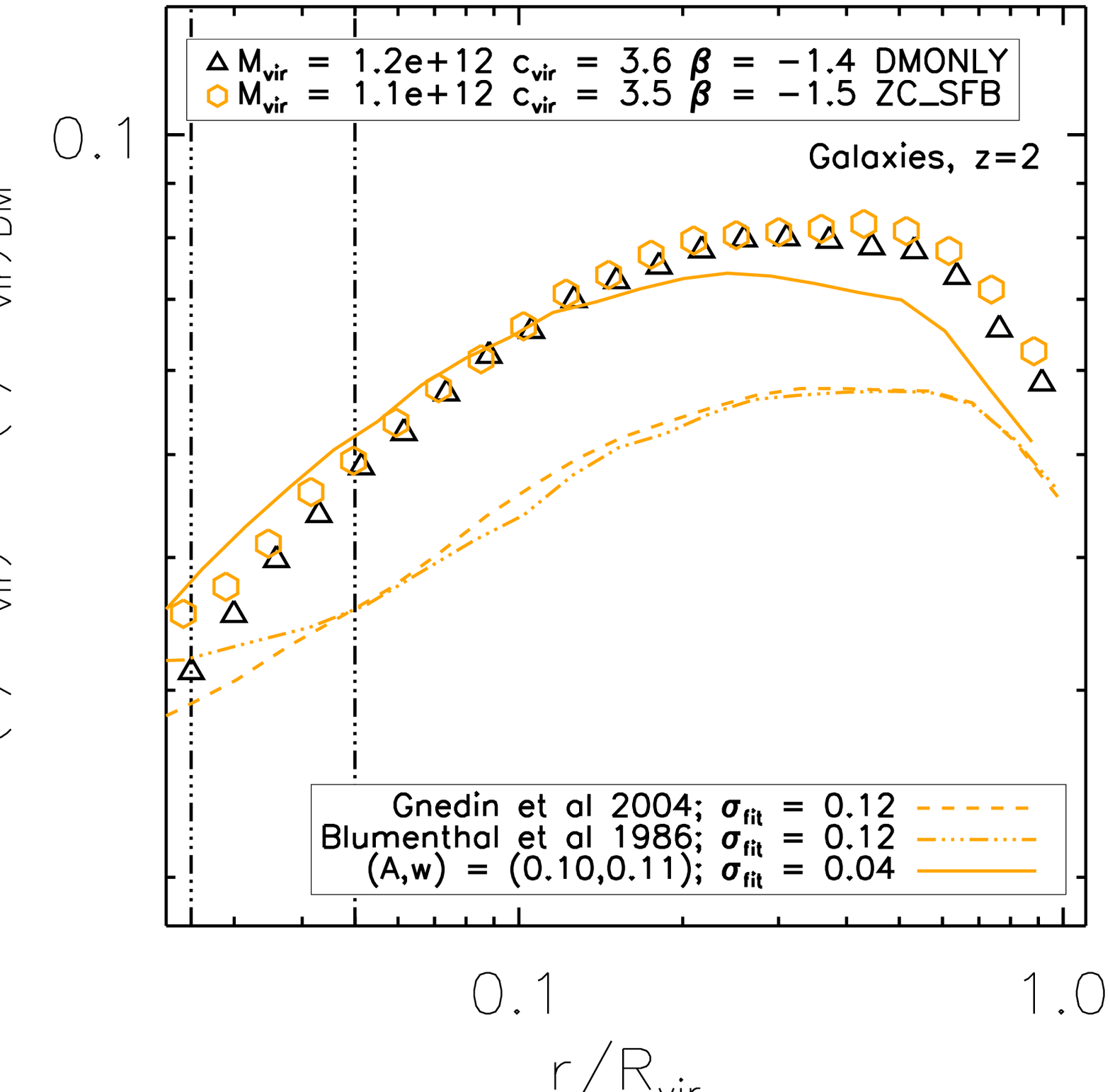, scale=0.3} \\
  \epsfig{figure=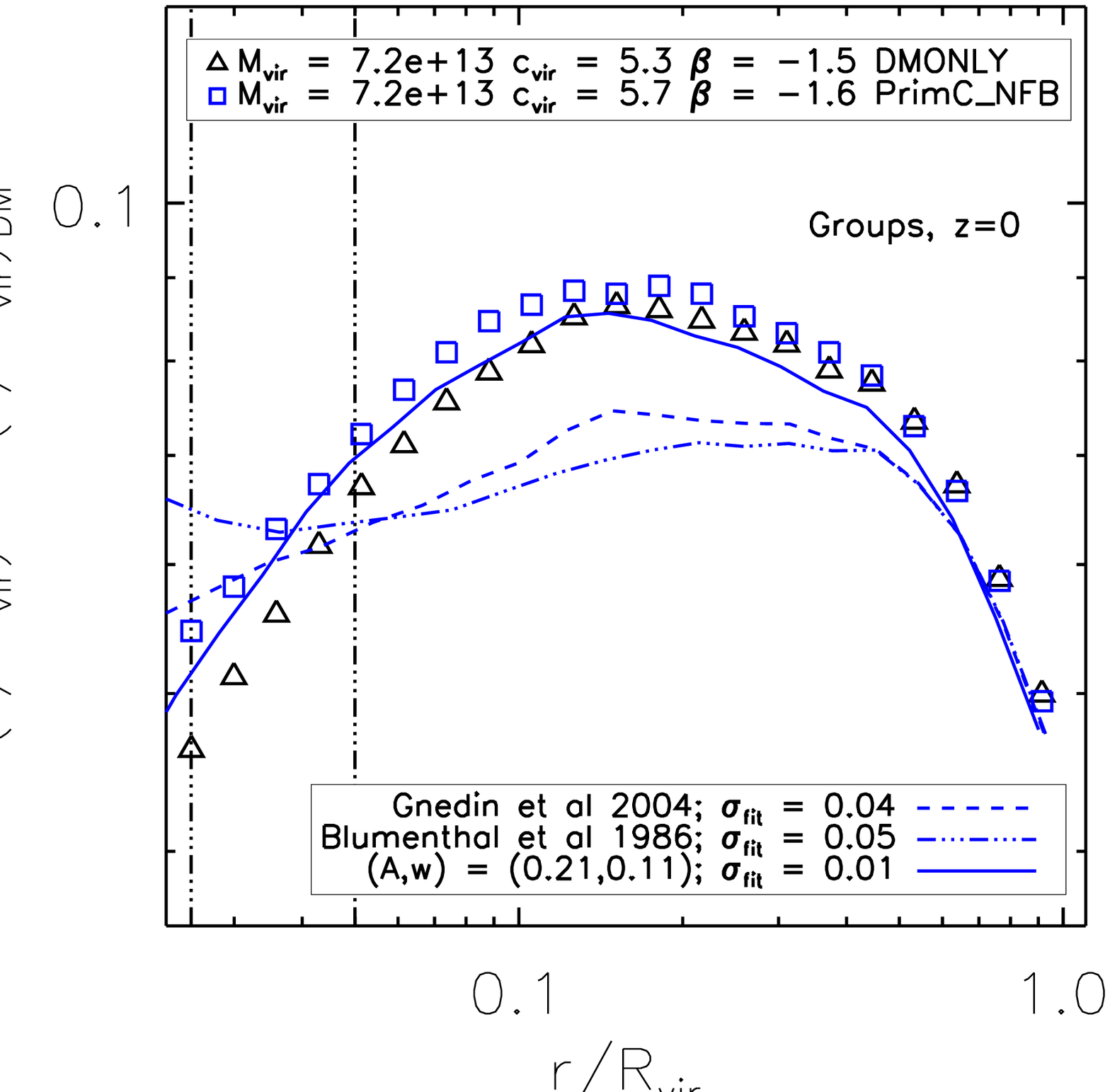, scale=0.3} &
  \epsfig{figure=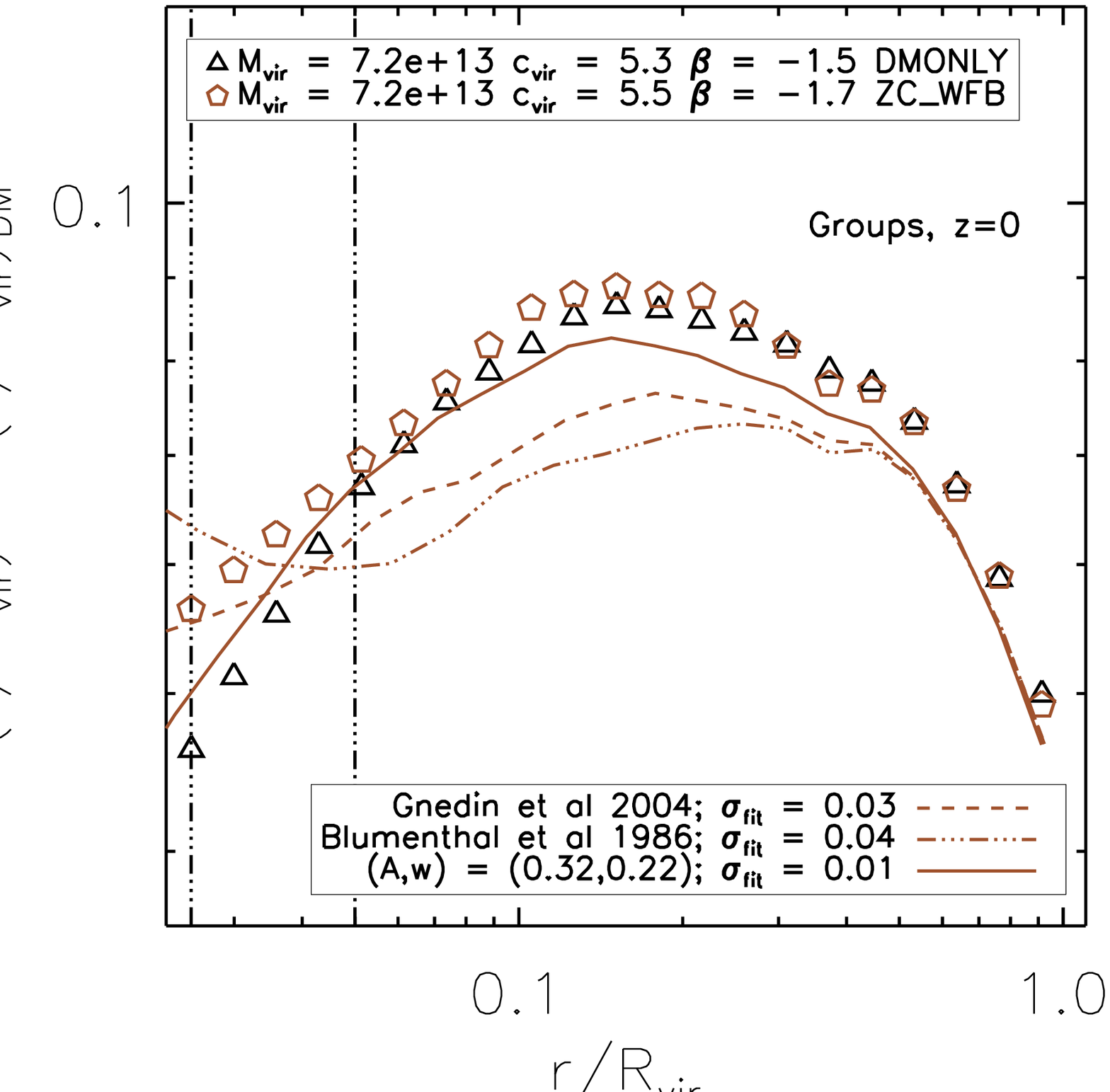, scale=0.3} &
  \epsfig{figure=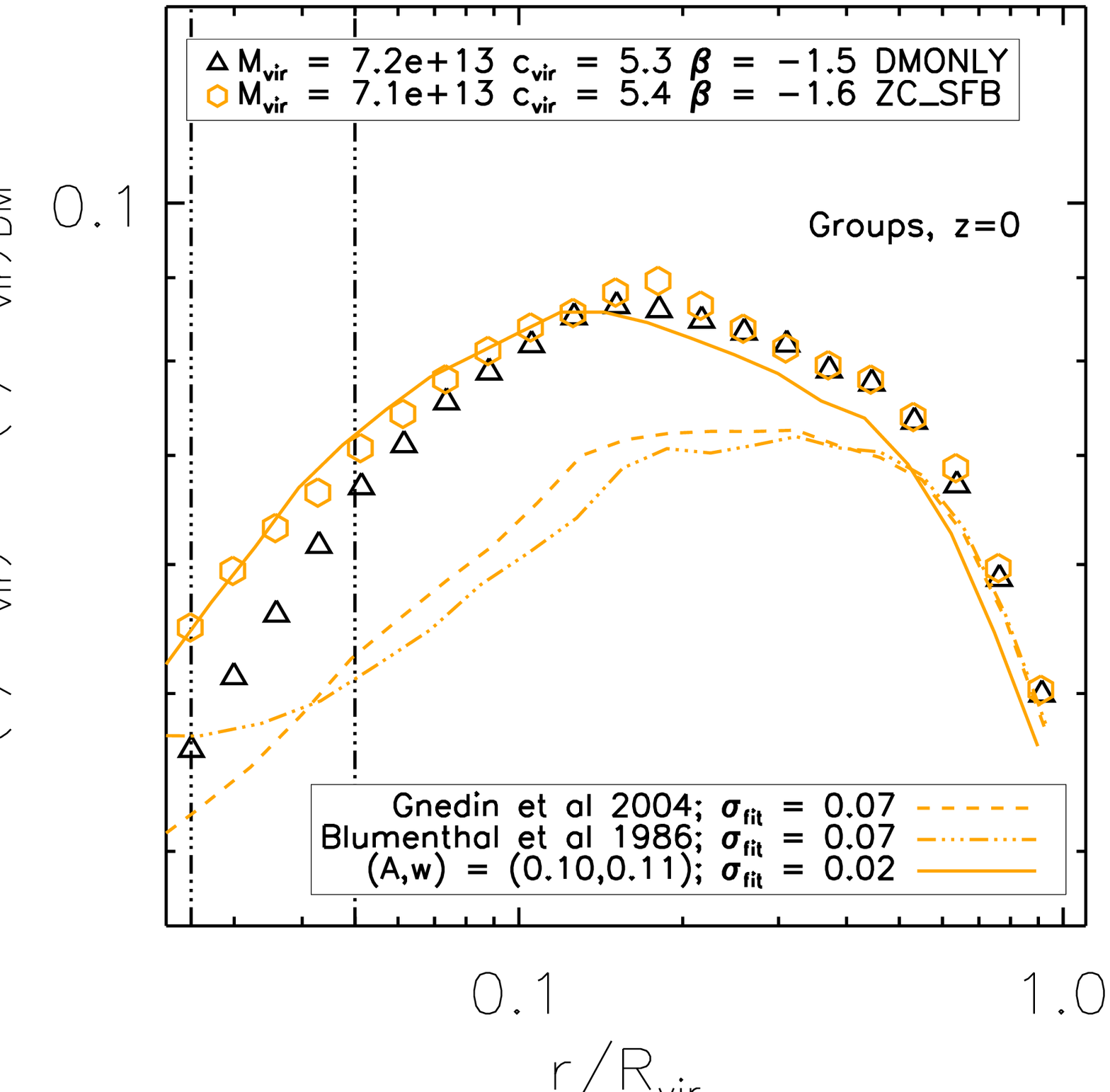, scale=0.3} \\
  \end{tabular}
  \caption{We test adiabatic contraction models by comparing the predicted mean DM density profiles, in average virial units 
    of the matched DMONLY haloes, for the simulations PrimC\_NFB (left), ZC\_WFB (middle)
    and ZC\_SFB (right) respectively. Also shown are the mean DMONLY
    profiles (which have been reduced by $(1-f^{\rm univ}_{\rm b})$ for comparison). 
    Overlaid are the predictions from
    the adiabatic contraction models of B86 (triple-dot-dashed), G04 using 
    their default parameter values (scaled to our definition of the virial radius)
    (dashed), and G04 with our best-fit parameter values (solid) as
    determined separately for each model, mass range, and redshift in
    Fig.~\ref{fig:avsw} along with the goodness-of-fit measure $\sigma_{\rm fit}$.
    Additionally, in the top legend we give the mean virial mass of the haloes,  $M_{\rm vir}$, 
    the best-fit NFW concentration, $\rm c_{vir}$, and inner profile slope, $\beta$.   
    The top (bottom) row shows results for haloes at $z=2$ ($z=0$). No one parameter combination 
    can reproduce the range of DM haloes, indicating that the slowly cooling gas picture, on which 
    adiabatic contraction models are based, is not sufficient to model the behaviour of the DM in a 
    live simulation. In fact, ignoring adiabatic contraction altogether gives the best results for 
    $r > 0.1\,R_{\rm vir}$.}
  \label{fig:ACmodels}
\end{figure*}

We first compute the best-fitting parameters $A$ and $w$ for a subset of our 
simulated haloes with baryons to see if our simulations prefer a specific 
combination and, if so, how this compares with those found in previous work. 
We focus on 3 models: the runs with no feedback (PrimC\_NFB), weak 
stellar feedback (ZC\_WFB) and strong stellar feedback
(ZC\_SFB). 
For each simulation we 
matched the haloes to the corresponding objects in DMONLY, producing an average 
density profile at each redshift. 
At $z=2$ we consider haloes with $M_{\rm vir}=[5-50]\times 10^{11}\hMsol$, while
at $z=0$ we study the range  $M_{\rm vir}=[3-60]\times 10^{13}\hMsol$. 

In Fig.~\ref{fig:avsw} we show the distribution of $\sigma_{\rm fit}$
values in the $A-w$ plane, for each of the
three simulation models at $z=2$ (top panels) and $z=0$ (bottom
panels). We calculate $\sigma_{\rm fit}$ using
equation~(\ref{eqn:sigmafit}), replacing $\rho_{\rm NFW}$ with the
appropriate adiabatically contracted density profile, over the range 
$0.025 \le r/R_{\rm vir} \le 1$.

The best-fit values of $A$ and $w$ depend strongly on
both halo mass and on the baryonic physics.
There is significant degeneracy between the parameters $A$ and $w$,
higher values of $A$ can be compensated by higher values of $w$. The
parameters suggested by G04\footnote{Note that we have rescaled their
  value of $A$ to our definition of the virial radius.} and the
original model by B86 work best 
for simulation ZC\_WFB, but even here they differ substantially from
our best-fit values. These findings are in good agreement with
\citet{Gustafsson:06}. 

\subsection{Predicted DM density profiles}

The DM density profiles predicted by the adiabatic contraction models are
shown in Fig.~\ref{fig:ACmodels}. When the parameter values
suggested by G04 are used, the G04 model predicts similar profiles to
B86, which do not describe the contracted DM
profiles well. The models typically underestimate the DM density for
$r \ga 10^{-1} R_{\rm vir}$ and more so for the simulations with
stronger feedback. If, on the other hand, we use the best-fit values of $A$ and $w$ for
each simulation and halo sample then the predictions of the G04 method
agree much better with the simulated profiles, but even in that case
one would obtain a closer match to the actual density profile by 
neglecting adiabatic contraction for $r> 0.1 R_{\rm vir}$. 

It is perhaps not too surprising that the models for adiabatic
contraction do not describe the simulations well. The assumption that
the baryons initially trace the DM halo profiles is clearly violated in hierarchical
models. Haloes are built by mergers of smaller progenitors, and cooling
and feedback have already redistributed the baryons in these
objects. Rather than contracting 
slowly as the gas cools, a large fraction of the baryons simply fall
in cold~\citep{Kay:00,Keres:05}. Moreover, in the stronger feedback models a substantial
fraction of the baryons is ejected. 
\citet{Tissera:09} recently demonstrated that the contraction
is manifestly not adiabatic. They find that the pseudo-phase space density relation
is strongly modified when baryons were added to high resolution DM only simulations.

Models for adiabatic contraction are required when full hydrodynamic
simulations are not available. Unfortunately, it is not possible to
predict what values of $A$ and $w$ to use for the G04 model without a
much better understanding of the baryonic physics. Moreover, even if
the physics were known, we would need to simulate many haloes because
the best-fit values of the parameters depend on both halo mass and
redshift. 

Our results suggest that it is better to ignore adiabatic contraction
for $r>0.1\,R_{\rm vir}$, but that the use of adiabatic contraction models such as those by B86
and G04 does typically represent an improvement at smaller radii, provided the feedback
is moderate or weak. 

\section{Conclusions}
\label{sec:conclusion}

Our main aim in this work was to investigate the response of the DM halo to 
the presence of baryons at a variety of masses and redshifts. 
We utilised a series of high-resolution simulations within a cosmological
volume, with a number of different prescriptions for the sub-grid physics,
to probe the effect of baryons on the DM distribution of haloes. Our results centred
on galaxy-scale haloes at $z=2$ and groups and clusters at $z=0$. 
We were particularly interested to discern the effect of the baryons when going from 
the situation in which radiative cooling dominates (leading to a high central concentration
of baryons in the form of stars and cold gas) to one where feedback dominates (reducing
the central galaxy mass and expelling gas from the halo). As we showed
in Fig.~\ref{fig:fbfbvir}, when comparing central and global baryon fractions,
our simulations with strong feedback (ZC\_WFB\_AGN and ZC\_SFB) 
reduce the baryon fraction $f_{\rm b}$, in comparison with PrimC\_NFB (no feedback), by factors of 2-3 
for {\it Galaxy} haloes at $z=2$. In {\it Group} and {\it Clusters} the AGN can remove 
nearly half of the baryons from the inner region of the halo, at $z=0$.

By comparing with observed stellar fractions in low redshift {\it Groups} and {\it Clusters} 
of galaxies
(Fig.~\ref{fig:fstar_m500}) we found that the simulations with a high baryon fraction 
(Fig.~\ref{fig:fbfbvir}) also predict stellar fractions significantly larger than observed. 
The simulation with inefficient gas cooling
and stellar feedback, PrimC\_WFB, and the strong feedback models, ZC\_WFB\_AGN and 
ZC\_SFB, are broadly in agreement with the observed stellar fractions in $z=0$ objects 
of mass $M_{\rm vir} \approx 10^{14}\,h^{-1}{\rm M_{\odot}}$. However, observed maximum star 
formation efficiencies of order 10\% - 20\% are only reproduced by the inclusion of AGN 
feedback.

However, these same strong feedback simulations are in disagreement with the
constraints inferred from combined gravitational lensing and stellar
dynamics analyses of the inner total mass density profile of massive,
early-type galaxies (Fig.~\ref{fig:totalbetafbprofile}). In this case
the efficient feedback prevents the steepening
of the density profile that is necessary to reproduce the observed
isothermal profiles. Instead, the simulations with high baryon
fractions are in closer agreement.
A more detailed comparison between the observations and simulations is
warranted, especially with regards to  
the biasing of lensing observations to steeper density profiles.

An enhanced baryon fraction in the inner halo is 
expected to contract the DM distribution, as a response to the deeper potential 
well of the system. This effect was clearly seen, from {\it Dwarf Galaxy} to {\it Cluster}
scales and at both low ($z=0$) and high ($z=2$) redshifts, and is summarised in
Fig.~\ref{fig:betafbprofile}, 
where haloes with larger central baryon fractions develop steeper central
profiles (especially the {\it Galaxy} haloes at high redshift). 
To quantify the contraction effect, we fit NFW profiles
and compared them to a simulation with no baryons. Variations in the concentration
are typically around 20 per cent or less, although the concentration of high-redshift 
dwarf galaxy haloes can be as much as 50 per cent higher when feedback effects are ignored. 
Strong feedback produces a mild \emph{decrease} in the concentration of a halo
through the removal of a significant amount of gas
from the halo, in a similar manner to what was found by~\citet{Koyama:08}. 

We also investigated non-parametric measures of concentration. We found that the ratio
$R_{500}/R_{2500}$ and the maximum circular velocity are both useful indicators
of the degree of contraction of the total mass profile. Efficient
feedback is required to redistribute the mass in the halo such that the maximum circular 
velocity is similar to the virial velocity, as is found observationally for disk galaxies. 

An interesting counterexample to the rule of increasing baryonic condensation leading to 
more concentrated DM haloes was witnessed. In our low redshift groups and clusters, the
dark matter is denser in the simulation with weak stellar feedback (PrimC\_WFB) 
than in the simulation with no feedback (PrimC\_NFB), even though the central baryon 
fractions are lower in the latter case.
A similar result was found by~\citet{Pedrosa:09a}, who analysed simulations of galaxies
with and without feedback. They argued
that the contraction of the DM halo is slowed by the infall of satellites (due to dynamical
friction). If the satellites are less-bound, they may be tidally disrupted before they
sink to the centre and their effect on halting the contraction is thus reduced. 

We compared our results to the adiabatic contraction models of 
\citet{Blumenthal:86} and the revised model of \citet{Gnedin:04}. We found that adiabatic contraction 
only improves the fits in the inner parts of the halo. Outside $0.1\,R_{\rm vir}$ one would do 
better \emph{not} to use any prescription at all.
Within $0.1\,R_{\rm vir}$ the former model is unable to reproduce the back-reaction on the DM. The latter model can do a reasonably good job, but only if its two 
parameters are allowed to vary with the baryonic physics, halo mass,
and redshift, in agreement with the findings of
\citet{Gustafsson:06}. This is of 
particular importance when feedback effects are strong, as required to reproduce the
cooled baryon fractions in low redshift groups and clusters. This therefore removes the 
predictive power of the model and we caution its use,
particularly if a detailed prediction for the structure of a DM halo is required. 

If one wishes to match the stellar fractions at low redshift, then models with strong feedback are required to suppress star formation. The total mass profiles in such simulations
are very similar to the DMONLY case, with $v_{\rm max} \approx V_{\rm
  vir}$, as observed. 
Intriguingly, the NFW concentration in these systems is actually \emph{reduced} relative to the 
same halo in the DMONLY run. 
The finding reported in \citet{Duffy:08b} that observed {\it groups} appear
to be significantly more concentrated than simulated haloes, is therefore still
unexplained. The hypothesis that the inclusion of baryons would
resolve the discrepancy  
between theory and observation has been shown to be wrong. Worse, the
disagreement actually 
grows larger if one utilises strong feedback physics schemes that can
reproduce the observed stellar fractions in these systems.

\section*{Acknowledgements}

The authors acknowledge the use of {\tt contra}, a public subroutine
generously supplied by Oleg Gnedin, for the calculation of both the G04 and
B86 adiabatic contraction models. Additional thanks to Volker Springel for
both comments and the use of {\tt SubFind} and {\tt Gadget}. We further thank
Ben Oppenheimer for his helpful comments.
ARD is supported by an STFC studentship and gratefully acknowledges
that this work was partly funded by both a Marie Curie ETS grant and
the European Science Foundation Short Visit fund, AstroSim. The
simulations presented here were run on Stella, the LOFAR 
BlueGene/L system in Groningen and on the Cosmology Machine at the
Institute for Computational Cosmology in Durham as part of the Virgo
Consortium research programme. This work
was sponsored by the National Computing Facilities Foundation (NCF) for
the use of supercomputer facilities, with financial support from the
Netherlands Organization for Scientific Research (NWO). 
This work was supported by Marie Curie Excellence Grant
MEXT-CT-2004-014112 and by an NWO VIDI grant.

\appendix
\section{Resolution Tests}
\label{appendix:restest}
To assess the effects of resolution on our results, we compare the
$512^3$ DMONLY and ZC\_WFB simulations with lower resolution versions
($256^3$ and $128^3$, with softening lengths increased by factors of 2 and 4 
respectively). 
We initially select haloes with $10^3$
DM particles from the lowest resolution simulation. Matching haloes in the higher
resolution simulations are found by first identifying high-resolution haloes that lie
within the virial radius, $R^{128}_{\rm pot}$, of the low resolution halo. 
We consider a halo to match if the candidate object fulfils
the following criteria
\begin{equation}
\frac{M_{\rm vir} - M^{128}_{\rm vir}}{M^{128}_{\rm vir}} \le 0.2 \,;\nonumber \\
\frac{R_{\rm pot} - R^{128}_{\rm pot}}{R^{128}_{\rm vir}} \le 0.2 \,.
\end{equation}
This final selection removes $\sim 10$ per cent of the haloes from the sample, 
these objects are in the process of merging and hence would create artifacts in
the averaged density profiles. We therefore perform resolution tests on
57 (27) matched haloes from ZC\_WFB at $z = 0$ ($2$) and 51 (27) haloes from
DMONLY at $z=0$ ($2$).
Note that this matching is different to the scheme employed in the rest of this work
which was based on linking identical DM particles between different physics simulations.

\begin{figure*}
  \begin{center}
    \begin{tabular}{cc}
    \epsfig{figure=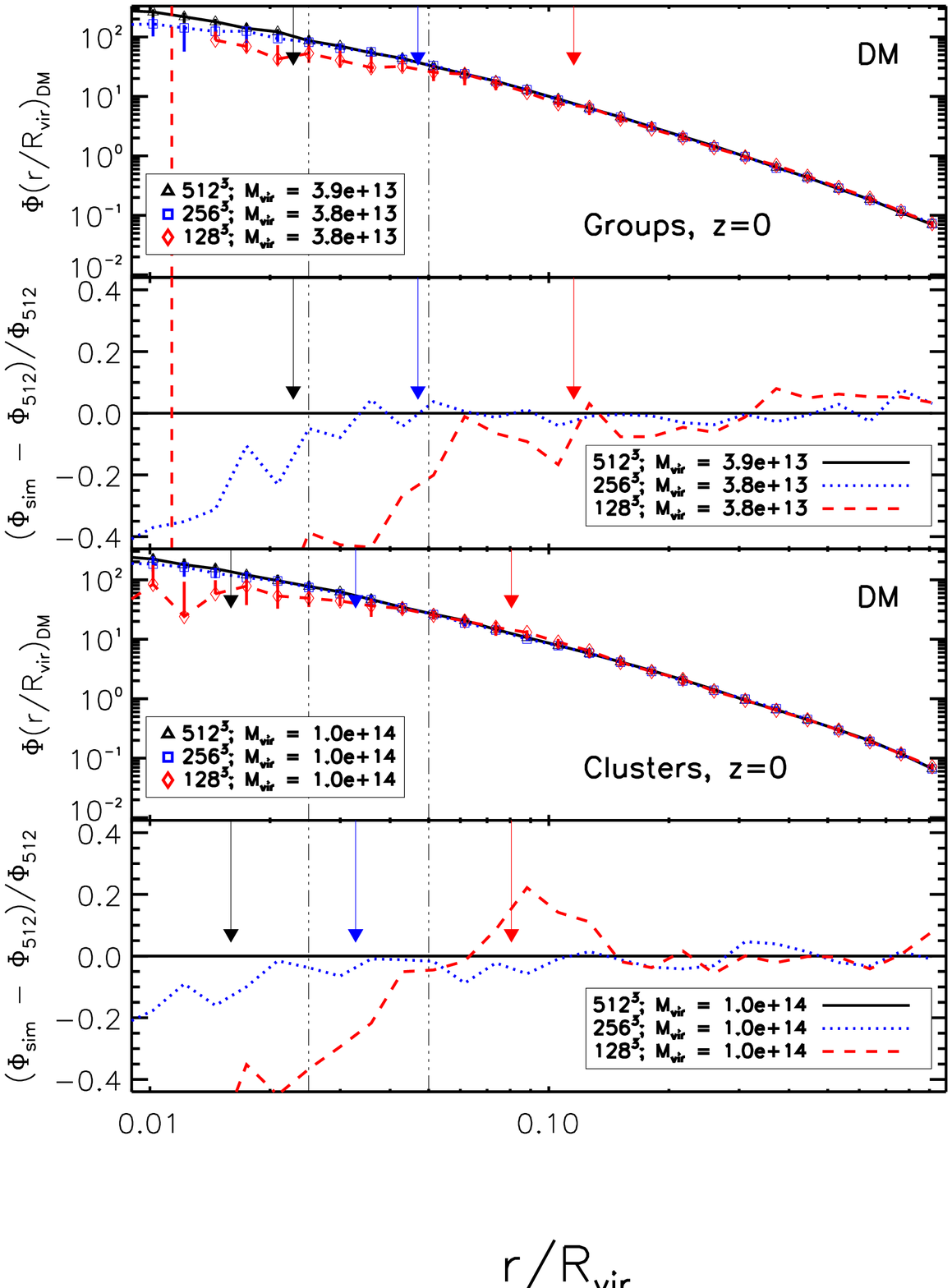, scale=0.35} & 
    \epsfig{figure=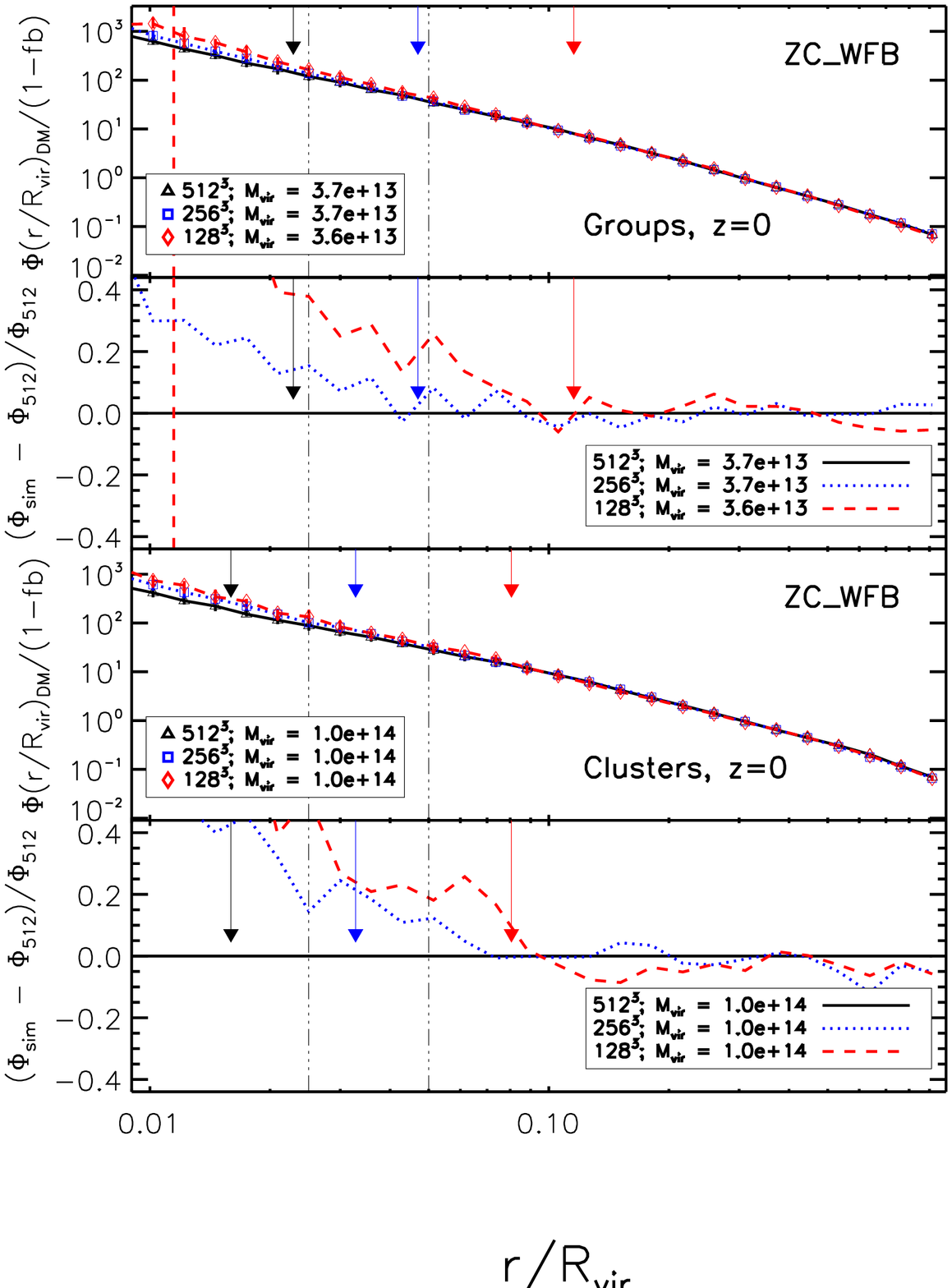, scale=0.35} \\
    \end{tabular}
    \caption{Comparison of mean DM density profiles (and their residuals) for {\it Group} and
    {\it Cluster} haloes (top and bottom panels respectively), drawn from the  $100\hMpc$ box at 
    $z=0$, for runs with different resolutions. Results from DMONLY are shown in the left column, 
    while results from ZC\_WFB are shown in the right column. 
    The colours indicate the resolution, from $128^3$ (red), to
    $256^3$ (blue) to $512^3$ (black). Most softening lengths are sufficiently small to fall outside
    of the plotted area but, where visible, softening scales are indicated with vertical lines with
    colour and line-style indicating simulation resolutions, described in the legend.
    The arrows indicate the P03 convergence radius for each simulation, according to colour.
    The vertical lines denote 2.5 and 5 per cent of the virial radius, corresponding to the scale 
    within which the inner density profile slope is measured. The legend contains the mean virial 
    mass of the haloes,  $M_{\rm vir}$.}
    \label{fig:conv_z0}
  \end{center}
\end{figure*}

\begin{figure*}
  \begin{center}
    \begin{tabular}{cc}
    \epsfig{figure=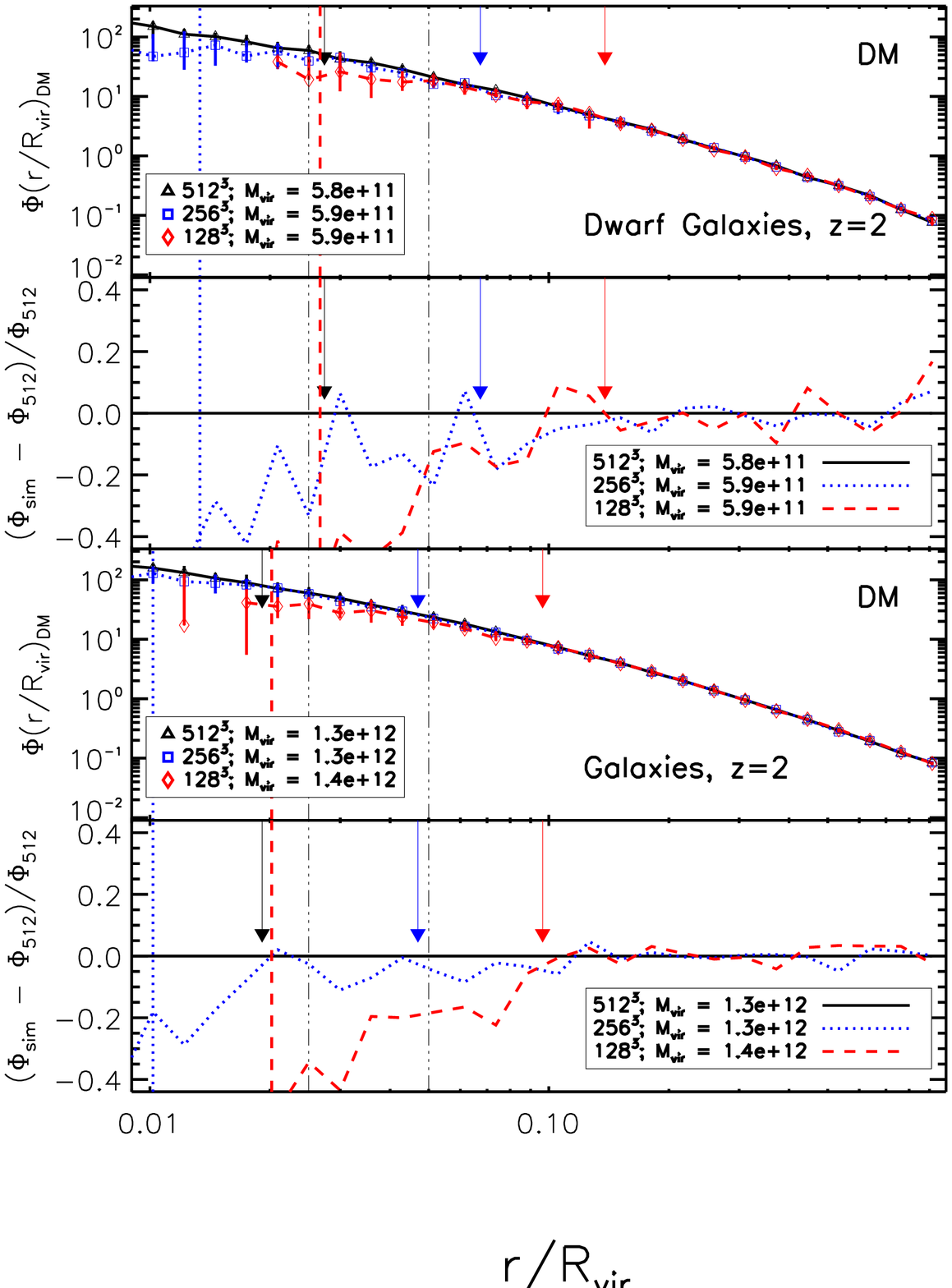, scale=0.35} & 
    \epsfig{figure=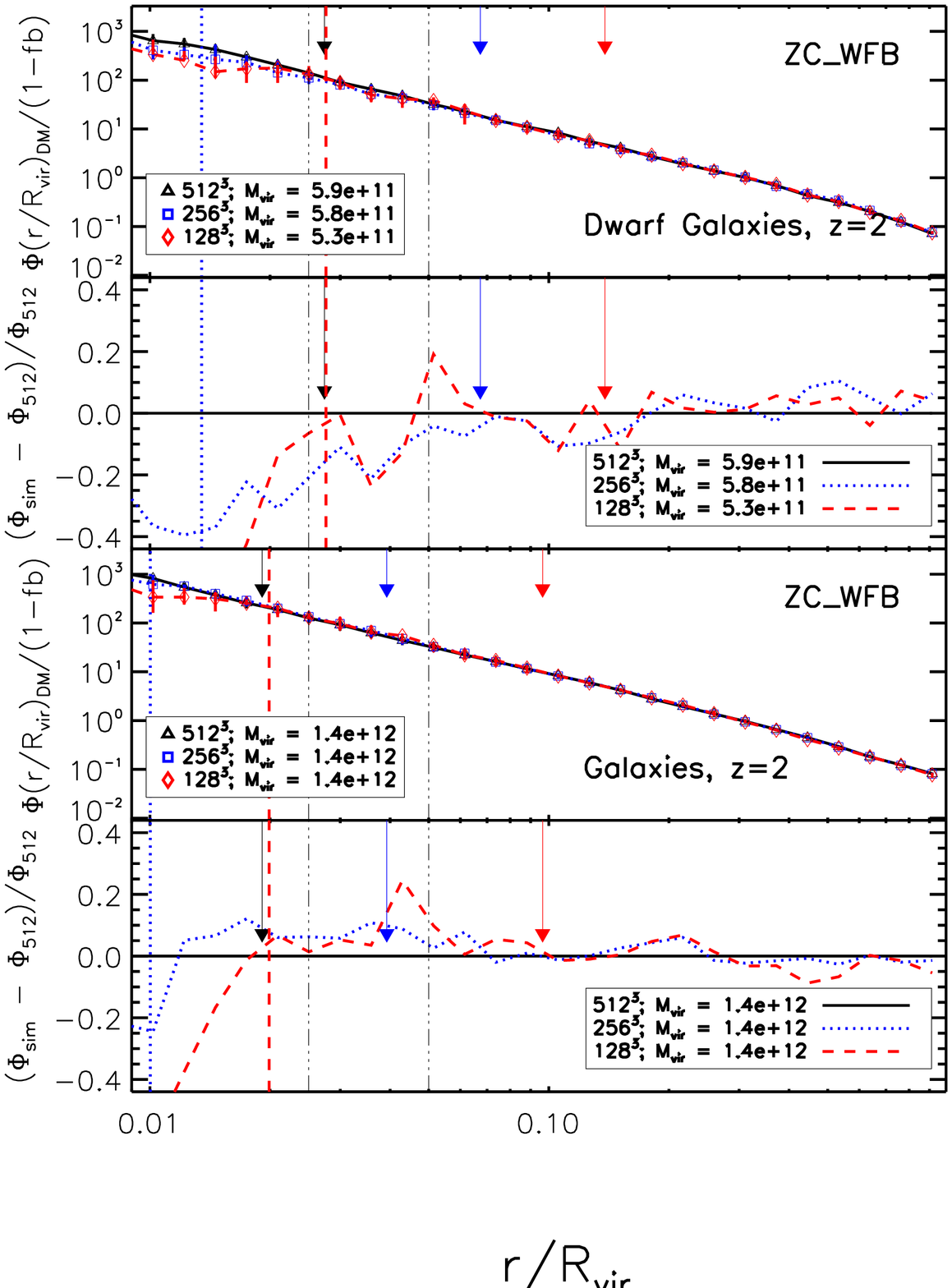, scale=0.35} \\
    \end{tabular}
    \caption{As in Fig.~\ref{fig:conv_z0}, but for {\it Dwarf Galaxy} (top panels) and {\it Galaxy} 
    (bottom panels) haloes at $z=2$, drawn from the $25 \hMpc$ box.}
    \label{fig:conv_z2}
  \end{center}
\end{figure*}

Fig.~\ref{fig:conv_z0} compares average total mass density 
profiles for a group and a cluster-scale halo at $z=0$ in the three runs. 
In the DMONLY case, resolution effects cause the density to be underestimated, 
whereas the density is overestimated for ZC\_WFB. At $z=2$, the results from 
(lower mass) dwarf and galaxy-scale haloes are similar, although the
density is also underestimated in the under-resolved regions in ZC\_WFB,
(Fig.~\ref{fig:conv_z2}). 

\citet{Power:03} defined a convergence radius, such that the mean two-body
relaxation time of the particles within that radius is of the order a Hubble time.
The equivalent radius is shown for each of our simulations in the two figures, as
a solid vertical arrow. Although this radius was originally applied to collisionless
$N$-body simulations (like DMONLY), we can see that it also provides a satisfactory
indication of where the density profile becomes numerically resolved
if baryons are included. This is especially visible in the lower sub-panels, where the fractional
deviations of the two lower-resolution profiles from our standard-resolution ($512^3$) 
results are shown.

\label{lastpage}
\end{document}